\documentclass[conference]{IEEEtran}

\newtheorem{definition}{Definition}

\newtheorem{lemma}{Lemma}
\newtheorem{corollary}{Corollary}
\newtheorem{theorem}{Theorem}
\newtheorem{example}{Example}

\usepackage{url}
\usepackage{amsfonts}
\usepackage{amssymb}
\usepackage{amsmath}
\usepackage{stmaryrd}
\usepackage{verbatim}
\usepackage{times}

\usepackage{todonotes}
\usepackage{etex}
\usepackage{booktabs}
\usepackage{tabularx}
\usepackage{multirow}
\usepackage{pgf}
\usepackage{pgfplots}
\usepackage{tikz}
\usetikzlibrary{arrows,calc,shapes}
\tikzset{p0node/.style={draw,rectangle,minimum width=7mm,minimum height=7mm,inner sep=0}}
\tikzset{p1node/.style={draw,regular polygon,regular polygon sides=3,minimum width=10mm,minimum height=10mm,inner sep=0}}
\tikzset{prnode/.style={draw,circle,minimum width=7mm,inner sep=0}}
\tikzset{panode/.style={draw,regular polygon,regular polygon sides=8,minimum width=10mm,minimum height=10mm,inner sep=0}}
\tikzset{pbnode/.style={draw,regular polygon,regular polygon sides=5,minimum width=10mm,minimum height=10mm,inner sep=0}}

\usepackage{xspace}
\usepackage{wrapfig}
\usepackage{cite}
\usepackage{paralist}
\usepackage{algorithm}
\usepackage{algpseudocode}

\colorlet{darkred}{red!80!black}
\colorlet{darkgreen}{green!70!black}
\colorlet{darkyellow}{yellow!90!black}

\renewcommand{\epsilon}{\varepsilon}

\newcommand{\prism}{\textsc{PRISM}\xspace}
\newcommand{\prismgames}{\textsc{PRISM-games}\xspace}
\newcommand{\llangle}{\langle\!\langle}
\newcommand{\rrangle}{\rangle\!\rangle}
\newcommand{\I}{\mathcal{I}}

\newcommand{\nats}{\mathbb{N}}
\newcommand{\rationals}{\mathbb{Q}}
\newcommand{\reals}{\mathbb{R}}

\newcommand{\strictposreals}{\reals^{> 0}}

\newcommand{\mg}{\mathcal{P}}

\newcommand{\vertexSet}{V}
\newcommand{\edgeSet}{E}
\newcommand{\colours}{\mathcal{C}}
\newcommand{\won}{\mathsf{won}}
\newcommand{\lost}{\mathsf{lost}}
\newcommand{\pri}{\mathsf{pri}}
\newcommand{\distr}{\mathit{Distr}}

\newcommand{\defeq}{\overset{\mathrm{def}}{=}}
\newcommand{\prob}{\mathsf{prob}}
\newcommand{\lprob}{\mathsf{lprob}}
\newcommand{\wprob}{\mathsf{wprob}}
\newcommand{\probm}{\mathbf{P}}
\newcommand{\win}{\mathsf{win}}
\newcommand{\plays}{\mathsf{Play}}
\newcommand{\strats}{\mathsf{Strats}}
\newcommand{\profit}{\mathsf{profit}}
\newcommand{\uprofit}{\mathsf{loss}}
\newcommand{\neutral}{\mathsf{neutral}}
\newcommand{\val}{\mathsf{val}}
\newcommand{\reach}{\mathsf{R}}
\newcommand{\arena}{\mathfrak{A}}
\newcommand{\winreg}{W}

\newcommand{\play}{\pi}

\newcommand{\functionDot}{\,\cdot\,}

\newcommand{\setcond}[2]{\{\, #1 \mid #2 \,\}}
\newcommand{\setnocond}[1]{\{#1\}}

\newcommand{\ltlPmax}{\mathcal{P}_{\!\mathrm{max}=?}}
\newcommand{\ltlU}{\mathbin{\mathbf{U}}}

\newcommand{\ltlF}{\mathord{\mathbf{F}}}

\iftrue
\newcommand{\at}[1]{{\color{teal}\textbf{AT:} #1}}
\newcommand{\mh}[1]{{\color{blue}\textbf{MH:} #1}}
\newcommand{\lz}[1]{{\color{purple}\textbf{LZ:} #1}}
\newcommand{\sven}[1]{{\color{red}\textbf{SS:} #1}}
\else
\newcommand{\at}[1]{}
\newcommand{\mh}[1]{}
\newcommand{\lz}[1]{}
\newcommand{\sven}[1]{}
\fi

\newcommand{\imc}{\textsc{IscasMC}}
\newcommand\qed{\hfill $\Box$}

\newenvironment{myproof}{\noindent\textit{Proof. }}{\nopagebreak
 \qed\medskip}

\newif\ifarxiv

\arxivtrue

\begin{document}

\title{Synthesising Strategy Improvement and Recursive Algorithms for Solving 2.5 Player Parity Games}

\iftrue
\author{
  \IEEEauthorblockN{Ernst Moritz Hahn\IEEEauthorrefmark{1}, Sven Schewe\IEEEauthorrefmark{2}, Andrea Turrini\IEEEauthorrefmark{1}, Lijun Zhang\IEEEauthorrefmark{1}}
  \IEEEauthorblockA{
    \IEEEauthorrefmark{1}State Key Laboratory of Computer Science, Institute of Software, Chinese Academy of Sciences, Beijing, China
  }
  \IEEEauthorblockA{
    \IEEEauthorrefmark{2}University of Liverpool, United Kingdom
}
  }
\fi

\maketitle

\begin{abstract}
2.5 player parity games combine the challenges posed by 2.5 player reachability games and the qualitative analysis of parity games.
These two types of problems are best approached with different types of algorithms: strategy improvement algorithms for 2.5 player reachability games and recursive algorithms for the qualitative analysis of parity games.
We present a method that---in contrast to existing techniques---tackles both aspects with the best suited approach and works exclusively on the 2.5 player game itself.
The resulting technique is powerful enough to handle games with several million states.
\ifarxiv
\else
\begin{center}
  Formal proofs are contained in the arXiv version at\\
  \url{http://TODO}
\end{center}
\fi
\end{abstract}

\section{Introduction}
Parity games are non-terminating zero sum games between two players, Player $0$ and Player $1$.
The players move a token along the edges of a finite graph without sinks.
The vertices are \emph{coloured}, i.e.\ labelled with a priority taken from the set of natural numbers.
The infinite sequence of vertices visited by the token is called the run of a graph, and each run is coloured according to the minimum priority that appears infinitely often on the run.
A run is winning for a player if the parity of its colour agrees with the parity of the player.

Parity games come in two flavours: games with random moves, also called 2.5 player games, and games without random moves, called 2 player games.
For 2 player games, the adversarial objectives of the two players are to ensure that the lowest priority that occurs infinitely often is even (for Player $0$) and odd (for Player $1$), respectively.
For 2.5 player games, the adversarial objectives of the two players are to maximise the likelihood that the lowest priority that occurs infinitely often is even resp.\ odd.

Solving parity games is the central and most expensive step in many model
checking~\cite{Kozen/83/mu,Emerson+all/93/mu,Wilke/01/Alternating,deAlfaro+Henzinger+Majumdar/01/Control,Alur+Henzinger+Kupferman/02/ATL},
satisfiability
checking~\cite{Wilke/01/Alternating,Kozen/83/mu,Vardi/98/2WayAutomata,Schewe+Finkbeiner/06/ATM},
and synthesis~\cite{Piterman/07/Parity,Schewe+Finkbeiner/06/Asynchronous}
methods.
As a result,
efficient algorithms for 2 player parity games have been studied intensively \cite{Kozen/83/mu,Emerson+Lei/86/Parity,Emerson+Jutla/91/Memoryless,McNaughton/93/Games,Browne-all/97/fixedpoint,Zielonka/98/Parity,Jurdzinski/00/ParityGames,Ludwig/95/random,Puri/95/simprove,Voge+Jurdzinski/00/simprove,BjorklundVorobyov/07/subexp,Obdrzalek/03/TreeWidth,Berwanger+all/06/ParityDAG,Jurdzinski/08/subex,Schewe/07/parity,Schewe/08/improvement,Fearnley/10/snare,ScheweTV15/symmetric}.

Parity games with 2.5 players have recently attracted attention \cite{DBLP:conf/stacs/ChatterjeeH06,ChatterjeeAH06,DBLP:conf/fossacs/Zielonka04,DBLP:journals/jcss/AlfaroM04,DBLP:conf/stoc/AlfaroM01,DBLP:conf/concur/ChatterjeeH06,DBLP:journals/jcss/ChatterjeeAH13,DBLP:journals/iandc/Chatterjee12,DBLP:journals/corr/abs-0712-1765}.
This attention, however, does not mean that results are similarly rich or similarly diverse as for 2 player games.
Results on the existence of pure strategies and on approximation algorithms \cite{DBLP:conf/fossacs/Zielonka04,DBLP:conf/stoc/AlfaroM01} are decades younger than similar results for 2 player games, while algorithmic solutions \cite{DBLP:conf/stacs/ChatterjeeH06,ChatterjeeAH06} focus on strategy improvement techniques only.

The qualitative counterpart of 2.5 player games, where one of the players has the goal to win almost surely while the other one wants to win with a non-zero chance, can be reduced to 2 player parity games, cf.~\cite{chatterjee-04-paper} or attacked directly on the 2.5 player game with recursive algorithm~\cite{HahnSTZ16}.
The more interesting quantitative analysis can be approached through a reduction to 2.5 player reachability games \cite{DBLP:conf/isaac/AnderssonM09}, which can then be attacked with strategy improvement algorithms \cite{Condon93onalgorithms,Ludwig/95/random,Puri/95/simprove,Fearnley/10/snare,ScheweTV15/symmetric}.
Alternatively, entangled strategy improvement algorithms can also run concurrently the 2.5 player parity game directly (for the quantitative aspects) and on a reduction to 2 player parity games (for the qualitative aspects) \cite{DBLP:conf/stacs/ChatterjeeH06,ChatterjeeAH06}. (Or, likewise, run on the larger game with an ordered quality measure that gives preference to the likelihood to win and uses the progress measure from \cite{BjorklundVorobyov/07/subexp} or \cite{Voge+Jurdzinski/00/simprove} as a tie-breaker.)

This raises the question if strategy improvement techniques can be directly applied on 2.5 player parity games, especially as such games are memoryless determined and therefore satisfy a main prerequisite for the use of strategy improvement algorithms.
The short answer is that strategy algorithms for 2.5 player parity games simply do not work.
Classical strategy improvement algorithms follow a joint pattern.
They start with an arbitrary strategy $f$ for one of the players (say Player $0$).
This strategy $f$ maps each vertex of Player $0$ to a successor, and thus resolves all moves of Player $0$.
This strategy is then \emph{improved} by changing the strategy $f$ at positions, where it is \emph{profitable} to do so.
The following steps are applied repeatedly until there is no improvement in Step 2.
\begin{enumerate}
 \item Evaluate the simpler game resulting from fixing $f$.
 \item Identify all changes to $f$ that, when applied once, lead to an improvement.
 \item Obtain a new strategy $f'$ from $f$ by selecting some subset of these changes.
\end{enumerate}

So where does this approach go wrong?
The first step works fine. After fixing a strategy for Player $0$, we obtain a 1.5 player parity game, which can be solved efficiently with standard techniques \cite{Courcoubetis+Yannakakis/95/Markov}.
It is also not problematic to identify the profitable switches in the second step.
The winning probability for the respective successor vertex provides a natural measure for the profitability of a switch.
We will show in Section \ref{sec:correctness} that, as usual for strategy improvement, any combination of such profitable switches will lead to an improvement.

The problem arises with the optimality guarantees.
Strategy improvement algorithms guarantee that a strategy that cannot be improved is optimal.
In the next paragraph, we will see an example, where this is not the case.
Moreover, we will see that it can be necessary to change several decisions in a strategy $f$ in order to obtain an improvement, something which is against the principles of strategy improvement.

\subsection{An illustrating example}
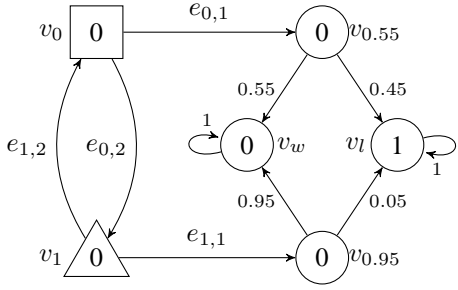
\begin{figure}
  \centering
  \begin{tikzpicture}[->,>=stealth',auto]
  \path[use as bounding box] (-1.25,-3.25) rectangle (4.8,0.4);
    \node[p0node] (p0) at (0,0) {0};
    \node at ($(p0)+(-0.6,0)$) {$v_{0}$};
    \node[p1node] (p1) at (0,-3) {0};
    \node at ($(p1)+(-0.6,0)$) {$v_{1}$};
    \node[prnode] (prob0) at (3,0) {0};
    \node at ($(prob0)+(0.7,0)$) {$v_{0.55}$};
    \node[prnode] (prob1) at (3,-3) {0};
    \node at ($(prob1)+(0.7,0)$) {$v_{0.95}$};
    \node[prnode] (ptarget) at (2,-1.5) {0};
    \node at ($(ptarget)+(0.6,0)$) {$v_{w}$};
    \node[prnode] (p2) at (4,-1.5) {1};
    \node at ($(p2)+(-0.55,0)$) {$v_{l}$};

    \draw (p0) to[bend left] node[left] {$e_{0,2}$} (p1);
    \draw (p1) to[bend left] node {$e_{1,2}$} (p0);
    \draw (p0) to node {$e_{0,1}$} (prob0);
    \draw (p1) to node {$e_{1,1}$} (prob1);
    \draw (prob0) to node[left] {\scriptsize$0.55$} (ptarget);
    \draw (prob0) to node[right] {\scriptsize$0.45$} (p2);
    \draw (prob1) to node[left] {\scriptsize$0.95$} (ptarget);
    \draw (prob1) to node[right] {\scriptsize$0.05$} (p2);
    \draw (ptarget) edge[loop left] node[above,very near end] {\scriptsize$1$} (ptarget);
    \draw (p2) edge[loop right] node[below,very near end] {\scriptsize$1$} (p2);
  \end{tikzpicture}
  \caption{A probabilistic parity game $\mg_{e}$.}
  \label{fig:example}
\end{figure}

  Consider the example 2.5 player parity game $\mg_{e}$ depicted in Figure~\ref{fig:example}.
  Square vertices are controlled by Player~$0$, while triangular ones are controlled by Player~$1$.
  In circular vertices, a random successor vertex is chosen with the given probability.
  In $v_{w}$, Player~$0$ wins with certainty (and therefore in particular almost surely), while she loses with certainty in $v_{l}$.
  In $v_{0.55}$ (or $v_{0.95}$), Player~$0$ wins with probability $0.55$ (or $0.95$).
For the nodes $v_{0}$ and $v_{1}$, we can see that the mutually \emph{optimal strategy} for Player~$0$ and Player~$1$ are to play $e_{0,2}$ and $e_{1,1}$, respectively.
  Player~$0$ therefore wins with probability $0.95$ when the game starts in $v_{0}$ and both players play optimally.

\subsection{Naive strategy iteration}
Strategy iteration algorithms start with an arbitrary strategy, and use an \emph{update rule} to get 
profitable switches These are edges, where the new target vertex has a higher probability of reaching the winning region (when applied once) compared to the current vertex.
As usual with strategy improvement, any combination of profitable switches leads to a strictly better strategy for Player $0$.
We illustrate that, if done naively, it may lead to values that are only locally maximal.
Assume that initially Player $0$ chooses the edge $e_{0,1}$ from $v_{0}$, then the best counter strategy of Player $1$ is to choose $e_{1,2}$ from $v_{1}$. 
The winning probability for Player $0$ under these strategies is $0.55$.

In strategy iteration, an update rule allows a player to switch actions only if the switching offers some \emph{improvement}. 
Since by switching to the edge $e_{0,2}$ Player $0$ would obtain the same winning probability, 
no strategy iteration can be applied, and the algorithm terminates with a sub-optimal solution.

Let us try to get some insights from this problem. Observe that Player~$1$ can entrap the play in the left vertices $v_{0}$ and $v_{1}$ when Player~$0$ chooses the edge $e_{0,2}$, such that the almost sure winning region of Player~$0$ cannot be reached.
However, this comes to the cost of losing almost surely for Player~$1$, as the dominating colour on the resulting run is $0$.
Broadly speaking, Player~$0$ must find a strategy that maximises her chance of reaching her almost sure winning regions, but only under the constraint that the counter strategy of Player~$1$ does not introduce new almost sure winning regions for Player~$0$.

\subsection{Solutions from the literature}
In the literature, two different solutions to this problem have been discussed.
Neither of these solutions works fully on the game graph of the 2.5 player parity game.
Instead, one of them uses a reduction to reachability games through a simple gadget construction \cite{DBLP:conf/isaac/AnderssonM09}, while the other uses strategy improvement on two levels, for the qualitative update described above, and for an update within subgames of states that have the same value \cite{ChatterjeeAH06,DBLP:conf/stacs/ChatterjeeH06};
this requires to keep a pair of entangled strategies.

\begin{figure}[b]
  \centering
  \begin{tikzpicture}[->,>=stealth',auto]
  \path[use as bounding box] (-0.5,-4.5) rectangle (6,0.3);
    \node[panode] (v) at (2,0) {$i$};
    \node at ($(v)+(0,-1)$) {$v$};
    \node[pbnode] (w) at (4,0) {$j$};
    \node at ($(w)+(0,-1)$) {$w$};
    \draw (v) -- (w);

    \node[panode] (vg) at (0,-3) {};
    \node at ($(vg)+(0,-1)$) {$v$};
    \node[prnode] (wgp) at ($(vg)+(2,0)$) {};
    \node at ($(wgp)+(-0.4,-0.4)$) {$w'$};
    \node[pbnode] (wg) at ($(vg)+(6,0)$) {};
    \node at ($(wg)+(0,-1)$) {$w$};

    \node[prnode] (won) at ($(wgp)+(0,1.2)$) {$\won$};
    \node[prnode] (lost) at ($(wgp)+(0,-1.2)$) {$\lost$};
    
    \draw (vg) -- (wgp);
    \draw (wgp) -- node {$1-\wprob-\lprob$}(wg);
    \draw (wgp) -- node {$\wprob$} (won);
    \draw (wgp) -- node {$\lprob$} (lost);
    
    \draw (won) to [out=375,in=345,looseness=8] node[right] {$1$} (won);
    \draw (lost) to [out=375,in=345,looseness=8] node[right] {$1$} (lost);
  \end{tikzpicture}
  \caption{Gadget construction.}
  \label{fig:gadget}
\end{figure}
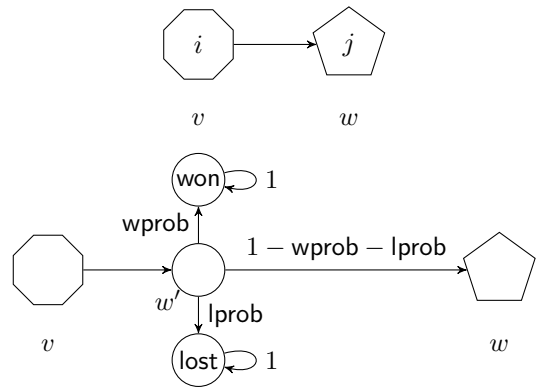

\paragraph*{Gadget construction for a reduction to reachability games}
In \cite{DBLP:conf/isaac/AnderssonM09}, it is shown that
2.5 player parity games can be solved by reducing them to 2.5 player reachability games and solving them, e.g.\ by using a strategy improvement approach.
For this reduction, one can use the simple gadgets shown in Figure~\ref{fig:gadget}.
There, when a vertex is passed by, the token goes to an accepting sink with probability $\wprob$ and to a losing sink with probability $\lprob$, both depending on the priority of the node (and continues otherwise as in the parity game).
For accordingly chosen $\wprob, \lprob$, any optimal strategy for this game is an optimal strategy for the parity game.
To get this guarantee, however, the termination probabilities have to be very small indeed.
In \cite{DBLP:conf/isaac/AnderssonM09}, they are constructed from the expression $(n!^2 2^{2n+3} M^{2n^2})^{-1}$ where $n$ is the number of vertices and $M$ is an integer depending on the probabilities occurring in the model.
Unfortunately, these small probabilities render this approach very inefficient and introduces numerical instability.

\paragraph*{Classic strategy improvement for 2.5 player parity games}
In \cite{DBLP:conf/stacs/ChatterjeeH06,ChatterjeeAH06}, the concept of strategy improvement algorithms has been extended to 2.5 player parity games.
To overcome the problem that the natural quality measure---the likelihood of winning---is not fine enough, this approach constructs classical 2 player games played on translations of the value classes (the set of vertices with the same likelihood of winning).
These subgames are translated using a gadget construction similar to the one used for qualitative solutions for 2.5 player to a solution to 2 player games from \cite{chatterjee-04-paper}.
This results in the 2 player game shown in Figure \ref{fig:quantexample}.

\begin{figure}
  \centering
  \resizebox{\linewidth}{!}{
  \begin{tikzpicture}[->,>=stealth',auto]
    \node[p0node] (p0) at (0,3) {0};
    \node[p1node] (p1) at ($(p0) + (0,-9)$) {0};
    \node[p1node] (p2) at ($(p1) + (5,0)$) {0};
    \node[p0node] (p3) at ($(p2) + (0,1.5)$) {0};
    \node[p1node] (p4) at ($(p3) + (0,1.5)$) {0};
    \node[p1node] (p5) at ($(p4) + (-1.5,1.5)$) {0};
    \node[p1node] (p6) at ($(p4) + (1.5,1.5)$) {1};
    \node[p1node] (p7) at ($(p4) + (0,3)$) {0};
    \node[p0node] (p8) at ($(p7) + (0,1.5)$) {0};
    \node[p1node] (p9) at ($(p8) + (0,1.5)$) {0};
    \node[p1node] (p10) at ($(p5) + (-1.25,0.85)$) {0};
    \node[p0node] (p11) at ($(p5) + (-1.25,-0.85)$) {0};
    \node[p1node] (p12) at ($(p6) + (0,1.5)$) {0};
    \node[p0node] (p13) at ($(p6) + (1.25,0.85)$) {1};
    \node[p0node] (p14) at ($(p6) + (1.25,-0.85)$) {1};
    \node[p0node] (p15) at ($(p6) + (0,-1.5)$) {1};

    \draw (p0) to[bend left=10] (p1);
    \draw (p1) to[bend left=10] (p0);
    \draw (p0) to (p9);
    \draw (p9) to (p8);
    \draw (p8) to (p7);
    \draw (p7) to (p6);
    \draw (p7) to (p5);
    \draw (p1) to (p2);
    \draw (p2) to (p3);
    \draw (p3) to (p4);
    \draw (p4) to (p5);
    \draw (p4) to (p6);
    \draw (p5) to (p11);
    \draw (p11) to (p10);
    \draw (p10) to (p5);
    \draw (p6) to (p13);
    \draw (p13) to (p12);
    \draw (p12) to (p6);
    \draw (p6) to (p14);
    \draw (p14) to (p15);
    \draw (p15) to (p6);
  \end{tikzpicture}
  }
  \caption{The qualitative game resulting from the game from Figure \ref{fig:example} when using the gadget construction from \cite{chatterjee-04-paper}.}
  \label{fig:quantexample}
\end{figure}

The strategy improvement algorithm keeps track of `witnesses $\omega = (\pi,\overline{\pi}_Q)$', which consists of a strategy $\pi$ on the 2.5 player parity game, and a strategy $\overline{\pi}_Q$ defined on the 2 player game $Q$ obtained from this 2.5 player game using the gadget construction from  \cite{chatterjee-04-paper}.
The strategies are entangled in that $\pi$ is the translation%
\footnote{In the notation of \cite{DBLP:conf/stacs/ChatterjeeH06,ChatterjeeAH06}, $\pi = \mathsf{Tr}_{\text{almost}}(\overline{\pi}_Q)$.}
of $\overline{\pi}_Q$.
That is, the strategies have to concur on the nodes of Player $0$ from the 2.5 player game, and each update on $\pi$ (resp.\ $\overline{\pi}_Q$) on the decisions from these vertices will translate to an update on the strategy of $\overline{\pi}_Q$ (resp.\ $\pi$) on the same vertices.

The valuation of one of these vertices is an ordered pair, consisting of the chance of obtaining the parity objective as the primary measure, and the value obtained in the quantitative game restricted to the individual value classes (vertices with the same chance of obtaining the parity objective) as a secondary measure \cite{DBLP:conf/stacs/ChatterjeeH06,ChatterjeeAH06}.

\subsection{Novel strategy iteration algorithm}
We show that
we can apply strategy improvement techniques with two different update rules directly on the 2.5 player game.
The first rule is a standard update rule for increasing the chance of reaching the almost sure winning region.
As we have seen in the example, this rule would not necessarily find the optimum:
it would not find the improvement from edge $e_{0,1}$ to $e_{0,2}$.
To overcome this problem, we introduce a second rule that handles the problem that Player~$1$ can reduce the chances of reaching the almost sure winning region of Player~$0$ by playing a strategy that leads to a larger almost sure winning region for Player~$0$.
This step uses a reduction to the \emph{qualitative} evaluation of these games.
Player $0$ changes her strategy in a way that she would win on the subgame that consists only of the edges of Player $0$ and Player $1$ that are \emph{neutral}.
For both players, these are the edges that lead to successor states with the same chance of winning under the current strategy.
If this provides a larger almost sure winning region for Player $0$ than $f$, then update $f$ in this \emph{new winning region} accordingly leads to a strictly better strategy $f'$.

While the first rule alone is not powerful enough, the two rules together provide the guarantee that a strategy that cannot be improved by either of them is optimal.

Note that the second rule is a non-standard rule for strategy improvement. Not only does it not rely on an improvement that is obtained when a change is applied once, it also requires to apply a fixed set of changes (in the new region) in one step for correctness. This is quite unusual for strategy improvement algorithms, where the combination of updates selected is irrelevant for correctness.

A further significant difference to the method from \cite{DBLP:conf/stacs/ChatterjeeH06,ChatterjeeAH06} is that we do not have to revert to solving transformed games.
Instead, we use the new generalisation of McNaughton's algorithm to the qualitative solution of 2.5 player parity games \cite{HahnSTZ16}.
This method seems to maintain the good practical performance known for classic recursive techniques, which have proven to be much faster than strategy improvement for the qualitative analysis of parity games \cite{FriedmannL09}.
A consequence of this choice is that we solve the qualitative games completely when there is no progress through the naive update step, which reduces the number of times that qualitative updates have to be considered.

This way, we use strategy improvement for the quantitative part of the analysis, where it has its strengths, while leaning on a variation \cite{HahnSTZ16} of McNaughton's algorithm \cite{Emerson+Lei/86/Parity,McNaughton/93/Games,Zielonka/98/Parity} for the qualitative part of the analysis, where prior research suggests that recursive algorithms outperform strategy improvement \cite{FriedmannL09}. 

Note that our quality measure strategy improvement is the same as the primary measure used in classical strategy improvement for 2.5 player parity games  \cite{DBLP:conf/stacs/ChatterjeeH06,ChatterjeeAH06}.
Different from that approach, we do not need to resort to gadget constructions for progressing within value classes, but can overcome the lack of progress w.r.t.\ the primary measure through invoking a performant algorithm for solving 2.5 player games quantitatively \cite{HahnSTZ16}.

\subsection{Organisation of the Paper}
We first introduce the standard terms and concepts in Section \ref{sec:prelim}.
We then recall the strategy improvement algorithms in Section~\ref{sec:strategyImprovement}, describe our algorithm in Section \ref{sec:algorithm}, show its correctness in Section \ref{sec:correctness}, and offer an experimental evaluation in Section \ref{sec:implementation}.

\section{Terms and Concepts}
\label{sec:prelim}

  A \emph{probability distribution} over a finite set $A$ is a function $\mu\colon A \to [0,1] \cap \rationals$ with $\sum_{a \in A} \mu(a) = 1$.
  By $\distr(A)$ we denote the set of probability distributions over $A$.

\begin{definition}
  \label{def:arena}
An \emph{arena} is a tuple $\arena = (\vertexSet_{0}, \vertexSet_{1}, \vertexSet_{r}, \edgeSet, \prob)$, where
\begin{itemize}
\item   $\vertexSet_{0}$, $\vertexSet_{1}$, and $\vertexSet_{r}$ are three finite disjoint sets of \emph{vertices} owned by the three players:
  Player~$0$, Player~$1$, and Player \emph{random}, respectively.
  Let $\vertexSet \defeq \vertexSet_{0} \cup \vertexSet_{1} \cup \vertexSet_{r}$;
\item
  $\edgeSet \subseteq \vertexSet \times \vertexSet$ is a set of \emph{edges} such that $(\vertexSet, \edgeSet)$ is a \emph{sinkless} directed graph, i.e.\ for each $v \in \vertexSet$ there exists $v' \in \vertexSet$ such that $(v,v') \in \edgeSet$;
  for $\sigma \in \setnocond{0, 1, r}$ we let $\edgeSet_{\sigma} \defeq \edgeSet \cap (\vertexSet_{\sigma} \times \vertexSet)$.
\item $\prob\colon \vertexSet_{r} \to \distr(\vertexSet)$ is the \emph{successor distribution function}.
  We require that for each $v \in \vertexSet_{r}$ and each $v' \in \vertexSet$, $\prob(v)(v') > 0$ if and only if $(v,v') \in \edgeSet$.
\end{itemize}
\end{definition}

If $\vertexSet_{0} = \emptyset$ or $\vertexSet_{1} = \emptyset$, we call $\arena$ a \emph{Markov decision process} (MDP) or 1.5 player game.
If both $\vertexSet_{0} = \vertexSet_{1} = \emptyset$, we call $\arena$ a \emph{Markov chain} (MC).
 Given an arena $\arena = (\vertexSet_{0}, \vertexSet_{1}, \vertexSet_{r}, \edgeSet, \prob)$, we define the following concepts.
  \begin{itemize}
  \item A \emph{play} is an infinite sequence  $\play = v_{0} v_{1} v_{2} v_{3} \ldots$ such that $(v_{i}, v_{i+1}) \in \edgeSet$ for all $i \in \nats$.
    We define $\play(i) \defeq v_i$.
    We denote by $\plays(\arena)$ the set of all plays of $\arena$.
  \item
    For $\sigma \in \setnocond{0,1}$, a (pure memoryless) \emph{strategy} $f_{\sigma}$ of Player~$\sigma$ is a mapping $f_{\sigma} \colon \vertexSet_{\sigma} \to \vertexSet$ from the vertices $\vertexSet_{\sigma}$ of Player~$\sigma$ to their successor states, i.e.\ for each $v \in \vertexSet_{\sigma}$, $(v, f_{\sigma}(v)) \in \edgeSet$.
    We denote the set of Player~$0$ and $1$ strategies by $\strats_{0}$ and $\strats_1$, respectively.
  \item Given a strategy $f_{0}$ for Player~$0$, we define the \emph{induced MDP} as
    $\arena_{f_{0}} = (\emptyset, \vertexSet_{1}, \vertexSet_{r} \cup \vertexSet_{0}, \edgeSet_{f_{0}}, \prob_{f_{0}})$ with $\edgeSet_{f_{0}} \defeq (\edgeSet \setminus \vertexSet_{0} \times \vertexSet) \cup \setcond{(v,f_{0}(v))}{v \in \vertexSet_{0}}$ and
	\[
		\prob_{f_{0}}(v)(v') {\defeq} \begin{cases}
			\prob(v)(v') & \text{if $v {\in} \vertexSet_{r}$,}\\
			1 & \text{if $v {\in} \vertexSet_{0}$ and $v' {=} f_{0}(v)$,}\\
			0 & \text{otherwise,}
		\end{cases}
	\]
    and similarly for Player~$1$.
  \item Given strategies $f_{0}, f_{1}$ for Player~$0$ and Player~$1$, respectively, we denote by $\arena_{f_{0},f_{1}} \defeq (\arena_{f_{0}})_{f_{1}}$ the \emph{induced MC} of the strategies.
 \item
  If $\arena$ is an MC, we denote by $\probm^{\arena}(v) \colon \Sigma^{\arena} \to [0,1]$ the uniquely induced~\cite{KemenySK66} probability measure on $\Sigma^{\arena}$,  the $\sigma$-algebra on the \emph{cylinder sets} of the plays of $\arena$, under the condition that the initial node is $v$.
    For general $\arena$, we let $\probm^{\arena}_{f_{0},f_{1}}(v) \defeq \probm^{\arena_{f_{0},f_{1}}}(v)$.
\end{itemize}

\begin{definition}
\label{def:game}
A \emph{2.5 player game}, also referred to as \emph{Markov game} (MG), is a tuple $\mg = (\vertexSet_{0}, \vertexSet_{1}, \vertexSet_{r}, \edgeSet, \prob, \win)$, where $\arena = (\vertexSet_{0}, \vertexSet_{1}, \vertexSet_{r}, \edgeSet, \prob)$ is an arena and
  $\win \subseteq \plays(\arena)$
  is the \emph{winning condition} 
  for Player~$0$, the set of plays for which Player~$0$ wins.
\end{definition}
The notions of plays, strategies, induced 1.5 player games, etc.\ extend to 2.5 player games by considering their underlying arena.

We consider two types of winning conditions, reachability and parity objectives.

\begin{definition}
  \label{def:quant_reachgame}
  A 2.5 player \emph{reachability game} is a 2.5 player game $\mg$ in which the winning condition $\win$ is defined by a \emph{target set} $\reach \subseteq \vertexSet$.
  Then, we have $\win = \setcond{\play \in \plays(\mg)}{ \exists i \geq 0: \play(i) \in \reach}$.
  For 2.5 player reachability games, we also use the notation $\mg = (\vertexSet_{0}, \vertexSet_{1}, \vertexSet_{r}, \edgeSet, \prob, \reach)$.
\end{definition}

\begin{definition}
\label{def:quant_paritygame}
A \emph{2.5 player parity game} (MPG) is a 2.5 player game $\mg$ in which the winning condition $\win$ is defined by the \emph{priority function} $\pri \colon \vertexSet \to \nats$ mapping each vertex to a natural number.
  We call the image of $\pri$ the set of \emph{priorities} (or: \emph{colours}), denoted by $\colours$.
  Note that, since $\vertexSet$ is finite, $\colours$ is finite as well.
  We extend $\pri$ to plays, using $\pri \colon \pi \mapsto \liminf_{i \to \infty} \pri(\play(i))$.
  Then, we have $\win = \setcond{\play \in \plays(\mg)}{\text{$\pri(\play)$ is even}}$.
  For 2.5 player parity games, we also use the notation $\mg = (\vertexSet_{0}, \vertexSet_{1}, \vertexSet_{r}, \edgeSet, \prob, \pri)$.
  We denote with $|\mg|$ the size of a 2.5 player parity game, referring to the space its overall representation takes.
\end{definition}

Note that in the above discussion we have defined strategies as mappings from vertices of the respective player to successor vertices.
More general definitions of strategies exist that e.g.\ use randomised choices (imposing a probability distributions over the edges chosen) or take the complete history of the game so far into account.
However, it is known that, for finite 2.5 player parity and reachability games, the simple pure memoryless strategies we have introduced above suffice to obtain mutually optimal infima and suprema~\cite{chatterjee-04-paper}.

We also use the common intersection and subtraction operations on directed graphs for arenas and games:
given an MG $\mg$ with arena $\arena = (\vertexSet_{0},\vertexSet_{1}, \vertexSet_{r}, \edgeSet, \prob)$,
\begin{itemize}
\item
	$\mg \cap \vertexSet'$ denotes the MG $\mg'$ we obtain when we restrict the arena $\arena$ to $\arena \cap \vertexSet' \defeq (\vertexSet_{0} \cap \vertexSet', \vertexSet_{1} \cap \vertexSet', \vertexSet_{r} \cap \vertexSet', \edgeSet \cap (\vertexSet' \times \vertexSet'), \prob{\restriction_{\vertexSet'\cap\vertexSet_{r}}})$,
\item for $\edgeSet' \supseteq \edgeSet_{r}$, we denote by $\mg \cap \edgeSet'$ the MG $\mg'$ we obtain when restricting arena $\arena$ to $\arena \cap \edgeSet' \defeq (\vertexSet_{0},\vertexSet_{1}, \vertexSet_{r}, \edgeSet \cap \edgeSet', \prob)$.
\end{itemize}
Note that the result of such an intersection may or may not be substochastic or contain sinks.
While we use these operations freely in intermediate constructions, we make sure that, whenever they are treated as games, they have no sinks and are not substochastic.

\begin{definition}
  \label{def:value}
Let $\mg = (\vertexSet_{0}, \vertexSet_{1}, \vertexSet_{r}, \edgeSet, \prob, \win)$ be a 2.5 player game, and let $f_{0}$ and $f_{1}$ 
two strategies for player $0$ and $1$ respectively. The \emph{value} $\val^{\mg}_{f_{0}, f_{1}} \colon \vertexSet \to [0,1]$ is defined as
  \[
  \val^{\mg}_{f_{0}, f_{1}}(v) \defeq \probm_{f_{0},f_{1}}^{\mg}(v)(\setcond{\play \in \plays(\mg)}{\play \in \win}) .
  \]
  We also define
\begin{align*}
\val^{\mg}_{f_{0}}(v) & \defeq \inf_{f'_{1} \in \strats_{1}} \val^\mg_{f_{0}, f'_{1}}(v)\text{,}\\
\val^{\mg}_{f_{1}}(v) & \defeq \sup_{f'_{0} \in \strats_{0}} \val^\mg_{f'_{0}, f_{1}}(v)\text{,} \\
\val^{\mg}(v) & \defeq \sup_{f'_{0} \in \strats_{0}} \inf_{f'_{1} \in \strats_{1}} \val^\mg_{f'_{0}, f'_{1}}(v)\text{.}
\end{align*}

We write $\val^{\mg}_{f'}\geq \val^\mg_f$ if, for all $v \in V$, $\val^{\mg}_{f'}(v)\geq \val^\mg_f(v)$ holds, and $\val^{\mg}_{f'}> \val^\mg_f$ if $\val^{\mg}_{f'}\geq \val^\mg_f$ and $\val^{\mg}_{f'}\neq \val^\mg_f$ hold.
\end{definition}

\begin{definition}
\label{def:winning}
  Given a vertex $v \in \vertexSet$, a strategy $f_{\sigma}$ for Player~$\sigma$ is called \emph{$v$-winning} if, starting from $v$, Player~$\sigma$ wins almost surely in the MDP defined by $f_{\sigma}$
  \pagebreak[3]
  (that is, $\val^\mg_{f_{\sigma}}(v)=1-\sigma$).
  For $\sigma \in \setnocond{0,1}$, a vertex $v$ in $\vertexSet$ is $v$-winning for Player~$\sigma$ if Player~$\sigma$ has a $v$-winning strategy $f_{\sigma}$.
  We call the set of $v$-winning vertices for Player~$\sigma$ the \emph{winning region} of Player~$\sigma$, denoted $\winreg_{\sigma}$. Note for $v\in\winreg_0$, $\val^\mg(v) = 1$, whereas for $v\in\winreg_1$
we have $\val^\mg(v) = 0$.
\end{definition}

\section{Strategy Improvement}
\label{sec:strategyImprovement}
A strategy improvement algorithm takes a memoryless strategy $f$ of one player, in our case of Player~$0$, and either infers that the strategy is optimal, or offers a family $\I_f$ of strategies, such that, for all strategies $f' \in \I_f$, $\val^{\mg}_{f'} > \val^{\mg}_{f}$ holds.

The family $\I_f$ is usually given through \emph{profitable switches}.
In such a case, $\I_{f}$ is defined as follows.

\begin{definition}
\label{def:profitable}
Given a 2.5 player game $\mg = (\vertexSet_{0}, \vertexSet_{1}, \vertexSet_{r}, \edgeSet, \prob, \win)$ and a strategy $f$ for Player~$0$, the \emph{profitable switches}, denoted $\profit(\mg,f)$, for Player~$0$ are the edges that offer a strictly higher chance of succeeding (under the given strategy).
That is, $\profit(\mg,f) = \setcond{(v,v') \in \edgeSet_{0}}{\val^{\mg}_{f}(v') > \val^{\mg}_{f}(v)}$.
We also define the unprofitable switches accordingly as $\uprofit(\mg,f) = \setcond{(v,v') \in \edgeSet_{0}}{\val^{\mg}_{f}(v') < \val^{\mg}_{f}(v)}$.

$\I_f$ is the set of strategies that can be obtained from $f$ by applying one or more profitable switches to $f$:
$\I_f = \setcond{f' \in \strats_{0}}{f' \neq f\text{ and }\forall v \in \vertexSet_{0}: f'(v) = f(v)\text{ or }(v,f'(v)) \in \profit(\mg,f)}$.
\end{definition}

Strategy improvement methods can usually start with an arbitrary strategy $f_0$, which is then updated by selecting some $f_{i+1} \in \I_{f_i}$ until $\I_{f_i}$ is eventually empty.
This $f_i$ is then guaranteed to be optimal.
The update policy with which the profitable switch or switches are selected is not relevant for the correctness of the method, although it does impact on the performance and complexity of the algorithms.
In our implementation, we use a `greedy switch all' update policy, that is we perform any switch we can perform and change the strategy to the locally optimal switch.

For 2.5 player reachability games, strategy improvement algorithms provide optimal strategies.

\begin{theorem}[cf.~\cite{Condon93onalgorithms}]
\label{theo:SI4reach}
For a 2.5 player reachability game $\mg$, a strategy improvement algorithm with the profitable switches / improved strategies as defined in Definition \ref{def:profitable} terminates with an optimal strategy for Player $0$.
\end{theorem}

In the strategy improvement step, for all $v \in V$ and all $f' \in \I_f$, it holds 
that $\val^\mg_{f'}(v) = \val^\mg_{f'}\big(f'(v)\big) \geq \val^\mg_f\big(f(v)\big) = \val^\mg_f\big(v\big)$. Moreover, strict inequality is obtained at some vertex in $V$.
 As we have seen in the introduction, this is not the case for 2.5 player parity games:
in the example from Figure \ref{fig:example}, for a strategy $f$ with $f(v_0)=v_{0.55}$, the switch from edge $e_{0,1}$ to $e_{0,2}$ is not profitable.
Note, however, that it is not unprofitable either.

\section{Algorithm}
\label{sec:algorithm}

We observe that situations where the naive strategy improvement algorithm described in the previous section gets stuck are \emph{tableaux}:
an improvement would be available, but among changes that are \emph{neutral} in that applying them once would neither lead to an increased nor to a decreased likelihood of winning.
As usual with strategy improvement algorithms, neutral switches cannot generally be added to the profitable switches: 
not only would one lose the guarantee to improve, one can also reduce the likelihood of winning when applying such changes.

Overcoming this problem is the main reason why strategy improvement techniques for MPG would currently have to use a reduction to 2.5 player reachability games (or other reductions), with the disadvantages discussed in the introduction.
We treat these tableaux directly and avoid reductions.
We first make formal what neutral edges are.

\begin{definition}
 Given a 2.5 player game $\mg = (\vertexSet_{0}, \vertexSet_{1}, \vertexSet_{r}, \edgeSet, \prob, \win)$ and a strategy $f$ for Player $0$, we define the set of \emph{neutral edge} $\neutral(\mg,f)$ as follows:
  \[
    \neutral(\mg,f) \defeq \edgeSet_{r} \cup \setcond{(v,v') \in \edgeSet_{0} \cup \edgeSet_{1}}{\val^{\mg}_{f}(v') = \val^{\mg}_{f}(v)}\text{.}
  \]
\end{definition}

Based on these neutral edges, we define an update policy on the subgame played only on the neutral edges.

\begin{definition}
\label{def:neutralSubgame}  
  Given a 2.5 player game $\mg = (\vertexSet_{0}, \vertexSet_{1}, \vertexSet_{r}, \edgeSet, \prob, \win)$ and a strategy $f$ for Player $0$, we define the \emph{neutral subgame} of $\mg$ for $f$ as
  $\mg' = \mg \cap \neutral(\mg,f)$.
  Based on $\mg'$ we define the set $\I_f'$ of \emph{additional strategy improvements} as follows.

  Let $\winreg_{0}$ and $\winreg_{0}'$ be the winning regions of Player $0$ on $\mg$ and $\mg'$, respectively.
  If $\winreg_{0} = \winreg_{0}'$, then $\I_f' = \emptyset$.
  Otherwise, let $\mathcal W$ be the set of strategies that are $v$-winning for Player $0$ on $\mg'$ for all vertices $v \in \winreg_{0}'$. Then we set
  \begin{align*}
	\I_f'' &  = \left\{f_0 \in \strats_{0}\;\middle|\kern-2pt\begin{array}{r}\exists f_w{\in} \mathcal W: \forall v {\in} \winreg_{0}': f_0(v) = f_w(v)\\\text{and } \forall v {\notin} \winreg_{0}': f_0(v)=f(v)\end{array}\kern-4pt\right\}\text{,}
  \\
	\I_f' & = \setcond{f' \in \I_f''}{\forall v \in W_0 : f'(v) = f(v)}\text{.}
  \end{align*}
\end{definition}
We remark that $\winreg_0\subseteq\winreg_0'$ always holds. Intuitively, we apply a qualitative analysis on the neutral subgame, and if the winning region of Player $0$ on the neutral subgame is larger than her winning region on the full game, then we use the new winning strategy on the new part of the winning region.
Intuitively, this forces Player $1$ to leave this area eventually (or to lose almost surely).
As he cannot do this through neutral edges, the new strategy for Player $0$ is superior over the old one.
\begin{example}
  Consider again the example MPG $\mg_{e}$ from Figure~\ref{fig:example} and the strategy such that $f_{0}(v_{0}) = v_{0.55}$.
Under this strategy, $\neutral(\mg_{e},f_{0}) = \edgeSet_{r} \cup \setnocond{(v_{0},v_{0.55}), (v_{0}, v_{1}), (v_{1},v_{0})}$;
the resulting neutral subgame $\mg'_{e}$ is the same as $\mg_{e}$ except for the edge $e_{1,1}$.
In $\mg'_{e}$, the winning region $\winreg'_{0}$ is $\winreg'_{0} = \setnocond{v_{0}, v_{1}, v_{w}}$, while the original region was $\winreg_{0} = \setnocond{v_{w}}$.
The two sets $\I'_{f_{0}}$ and $\I''_{f_{0}}$ contain only the strategy $f'_{0}$ such that $f'_{0}(v_{0}) = v_{1}$.
In order to avoid to lose almost surely in $\winreg'_{0}$, Player~$1$ has to change his strategy from $f_{1}(v_{1}) = v_{0}$ to $f'_{1}(v_{1}) = v_{0.95}$ in $\mg_{e}$.
Consequently, strategy $f'_{0}$ is superior to $f_{0}$: the resulting winning probability is not $0.55$ but $0.95$ for $v_0$ and~$v_1$.
\end{example}

Note that using $\I_f'$ or $\I_f''$ in the strategy iteration has the same effect. Once a run has reached $W_0$ in the neutral subgame, it cannot leave it.
Thus, changing the strategy $f_0$ from $\I_f''$ to a strategy $f'$ with $f'(v) = f(v)$ for $v \in W_0$ and $f'(v) = f_0(v)$ for $v \notin W_0$ will not change the chance of winning: $\val^{\mg'}_{f_0} = \val^{\mg'}_{f'}$ and  $\val^{\mg}_{f_0} = \val^{\mg}_{f'}$. 
This also implies $\I_f'' \neq \emptyset \Rightarrow \I_f' \neq \emptyset$, since $\I'_{f}$ contains all strategies that belong to $\I''_{f}$ and that agree with $f$ only on the original winning region $\winreg_{0}$.
Using $\I_f'$ simplifies the proof of Lemma \ref{lem:asImprove}, but it also emphasises that one does not need to re-calculate the strategy on a region that is already winning.

Our extended strategy improvement algorithm applies updates from either of these constructions until no further improvement is possible.
That is, we can start with an arbitrary Player $0$ strategy $f_0$ and then apply
$f_{i+1} \in \I_{f_i} \cup \I_{f_i}'$
until $\I_{f_i} = \I_{f_i}' = \emptyset$.
We will show that therefore $f_i$ is an optimal Player $0$ strategy.

For the algorithm, we need to calculate $\I_{f_i}$ and $\I_{f_i}'$.
Calculating $\I_{f_i}$ requires only to solve 1.5 player parity games \cite{Courcoubetis+Yannakakis/95/Markov}, and we use \imc~\cite{HLSTZ/14/IscasMC,HLSTZ/15/Lazy} to do so.
Calculating $\I_{f_i}'$ requires only qualitative solutions of neutral subgame $\mg'$.
For this, we apply the algorithm from \cite{HahnSTZ16}.

A more algorithmic representation of our algorithm with a number of minor design decisions is provided in
\ifarxiv
Appendix~\ref{apx:algorithm}.
\else
the arXiv version of this paper.
\fi
The main design decision is to favour improvements from $\I_{f_i}$ over those from $\I_{f_i}'$. This allows for calculating $\I_{f_i}'$ only if $\I_{f_i}$ is empty.
Starting with calculating $\I_{f_i}$ first is a design decision, which is slightly arbitrary.
We have made it because solving 1.5 player games quantitatively is cheaper than solving 2.5 player games qualitatively and we believe that the guidance for the search is, in practice, better in case of quantitative results.
Likewise, we have implemented a `greedy switch all' improvement strategy, simply because this is believed to behave well in practice.
We have, however, not collected evidence for either decision and acknowledge that finding a good update policy is an interesting line of future research.

\section{Correctness}
\label{sec:correctness}

\subsection{Correctness proof in a nutshell}
The correctness proof combines two arguments: the correctness of \emph{all} basic strategy improvement algorithms for reachability games and a reduction from 2.5 player parity games to 2.5 player reachability games with arbitrarily close winning probabilities for similar strategy pairs.
In a nutshell,
if we approximate close enough, then three properties hold for a game $\mg$ and a strategy $f$ of Player $0$:
\begin{enumerate}
 \item all `normal' strategy improvements of the parity game correspond to strategy improvements in the reachability game (Corollary~\ref{cor:profitable});
 \item if Player~$0$ has a larger winning region $W_0'$ in the neutral subgame (cf.\ Definition \ref{def:neutralSubgame}) for $P \cap \neutral(\mg,f)$ than for $\mg_f$, then replacing $f$ by a winning strategy in $\I_f'$ leads to an improved strategy in the reachability game (Lemma~\ref{lem:asImprove}); and
 \item if neither of these two types of strategy improvements are left, then a strategy improvement step on the related 2.5 player reachability game will not lead to a change in the winning probability on the 2.5 player parity game (Lemma~\ref{lem:noImprovement}).
\end{enumerate}

\subsection{Two game transformations}
In this subsection we discuss two game transformations that change the likelihood of winning only marginally and preserve the probability of winning, respectively.
The first transformation turns 2.5 player parity games into 2.5 player reachability games such that a strategy that is optimal strategy for the reachability game is also optimal for the parity game (cf.~\cite{DBLP:conf/isaac/AnderssonM09}).
\begin{definition}
  Let $\mg = (\vertexSet_{0}, \vertexSet_{1}, \vertexSet_{r}, \edgeSet, \prob, \pri)$, and 
let  $\epsilon \in (0,1)$ and $n \in \nats$.
We define the 2.5 player reachability game $\mg_{\epsilon,n} = (\vertexSet_{0}, \vertexSet_{1}, \vertexSet_{r}'', \edgeSet'', \prob',\setnocond{\won})$ with
\begin{itemize}
 \item $\vertexSet_r'' = \vertexSet_r \cup \vertexSet' \cup \setnocond{\won, \lost}$, where
(i) $\vertexSet'$ contains primed copies of the vertices; for ease of notation, the copy of a vertex $v$ is referred to as $v'$ in this construction;
 (ii)
 $\won$ and $\lost$ are fresh vertices; they are a winning and a losing sink, respectively;
 \item $\edgeSet' = \setcond{(v,w')}{(v,w) \in \edgeSet} \cup \setnocond{(\won,\won),(\lost,\lost)}$;
 \item $\edgeSet'' = \edgeSet' \cup \setcond{(v',v)}{v \in \vertexSet} \cup \setcond{(v',\won)}{v \in \vertexSet} \cup \setcond{(v',\lost)}{v \in \vertexSet}$;
 \item $\prob'(v)(w') = \prob(v)(w)$ for all $v \in \vertexSet_r$ and $(v,w) \in \edgeSet$;
 \item $\prob'(v')(\won) = \wprob\big(\epsilon,n,\pri(v)\big)$,
 \item $\prob'(v')(\lost) =  \lprob\big(\epsilon,n,\pri(v)\big)$,
 \item $\prob'(v')(v) = 1-\wprob\big(\epsilon,n,\pri(v)\big) - \lprob\big(\epsilon,n,\pri(v)\big)$ for all $v \in \vertexSet$, and
 \item $\prob'(\won)(\won) = \prob'(\lost)(\lost) = 1$.
\end{itemize}
where $\lprob,\wprob \colon (0,1) \times \nats \times \nats \rightarrow [0,1]$ are two functions with $\lprob(\epsilon,n,c)+\wprob(\epsilon,n,c) \leq 1$ for all $\epsilon \in (0,1)$ and $n,c \in \nats$.
\end{definition}
Intuitively, this translation replaces all the vertices by the gadgets from Figure~\ref{fig:gadget}.

Note that $\mg_{\epsilon,n}$ and $\mg$ have similar memoryless strategies.
By a slight abuse of the term, we say that a strategy $f_{\sigma}$ of Player~$\sigma$ on $\mg_{\epsilon,n}$ is \emph{similar} to her strategy $f'_{\sigma}$ on $\mg$ if $f'_{\sigma} \colon v \mapsto f_{\sigma}(v)'$ holds, i.e.\ when $v$ is mapped to $w$ by $f_{\sigma}$, then $v$ is mapped to $w'$ by $f'_{\sigma}$.

\begin{theorem}
[cf.~\cite{DBLP:conf/isaac/AnderssonM09}]
\label{theo:epsilon}
Let $\mg = (\vertexSet_{0}, \vertexSet_{1}, \vertexSet_{r}, \edgeSet, \prob, \pri)$ be a 2.5 player parity game. 
Then, there exists  $\epsilon \in (0,1)$,
 $n\geq |\mg| $ such that we can construct $\mg_{\epsilon,n}$ and the following holds:
For all strategies $f_{0}\in \strats_{0}$,  $f_{1}\in \strats_{1}$, and all vertices $v \in \vertexSet$, $\big|\val_{f_{0},f_{1}}^{\mg}(v) - \val_{f'_{0},f'_{1}}^{\mg_{\epsilon,n}}(v)\big| < \epsilon$, $\big|\val_{f_{0},f_{1}}^{\mg}(v) - \val_{f'_{0},f'_{1}}^{\mg_{\epsilon,n}}(v')\big| < \epsilon$, $\big|\val_{f_{0}}^{\mg}(v) - \val_{f'_{0}}^{\mg_{\epsilon,n}}(v)\big| < \epsilon$, $\big|\val_{f_{0}}^{\mg}(v) - \val_{f'_{0}}^{\mg_{\epsilon,n}}(v')\big| < \epsilon$, $\big|\val_{f_{1}}^{\mg}(v) - \val_{f'_{1}}^{\mg_{\epsilon,n}}(v)\big| < \epsilon$, and $\big|\val_{f_{1}}^{\mg}(v) - \val_{f'_{1}}^{\mg_{\epsilon,n}}(v')\big| < \epsilon$  holds,
where $f'_{0}$ resp.\ $f'_{1}$ are similar to $f_{0}$ resp.\ $f_{1}$.
\end{theorem}

The results of \cite{DBLP:conf/isaac/AnderssonM09} are stronger in that they show that the probabilities grow sufficiently slow for the reduction to be polynomial, but we use this construction only for correctness proofs and do not apply it in our algorithms. For this reason, existence is enough for our purpose.
As \cite{DBLP:conf/isaac/AnderssonM09} does not contain a theorem that directly makes the statement above, we have included a simple construction (without tractability claim) with a correctness proof in
\ifarxiv
Appendix \ref{apx:parity2reach}.
\else
the arXiv version of this paper.
\fi

We will now introduce a second transformation that allows us to consider changes in the strategies in many vertices at the same time.
\begin{definition}
Let $\mg = (\vertexSet_{0}, \vertexSet_{1}, \vertexSet_{r}, \edgeSet, \prob, \win)$ and a region $R \subseteq \vertexSet$. Let $\mathcal{F}_R = \setcond{f \colon R \cap \vertexSet_{0} \to \vertexSet}{\forall v \in R: \big(v,f(v) \big)\in \edgeSet}$ denote the set of memoryless strategies for Player~$0$ restricted to $R$.
The transformation results in a parity game $\mg^R = (\vertexSet_{0}', \vertexSet_{1}', \vertexSet_{r}', \edgeSet', \prob',\pri')$ such that
\begin{itemize}
\item $\vertexSet_{0}'' = \vertexSet_{0} \cup R$, $\vertexSet_{0}''' = (\vertexSet_{0} \cap R) \times \mathcal F_R$, and $\vertexSet_{0}' = \vertexSet_{0}'' \cup \vertexSet_{0}'''$;
 \item $\vertexSet_{1}'' = \vertexSet_{1} \setminus R$, $\vertexSet_{1}''' =  (\vertexSet_{1} \cap R) \times \mathcal F_R$, and $\vertexSet_{1}' = \vertexSet_{1}'' \cup \vertexSet_{1}'''$;
 \item $\vertexSet_r'' = \vertexSet_r \setminus R$, $\vertexSet_r''' =  (\vertexSet_{r} \cap R) \times \mathcal F_R$, and $\vertexSet_r' = \vertexSet_r'' \cup \vertexSet_r'''$;
 \item $\edgeSet' = \setcond{(v,w) \in \edgeSet}{v \in \vertexSet \setminus R} \cup \setcond{(v,(v,f))}{v \in R \text{ and } f \in \mathcal F_R} \cup \setcond{((v,f),(w,f))}{v,w \in R,\ (v,w) \in \edgeSet \text{ and either } v \notin \vertexSet_{0} \text{ or } f(v)=w} \cup \setcond{((v,f),w)}{v \in R,\ w \notin R,\ (v,w) \in \edgeSet \text{ and either } v \notin \vertexSet_{0} \text{ or } f(v)=w}$;
 \item $\prob'(v)(w) = \prob(v)(w)$, $\prob'\big((v,f)\big)(w) = \prob(v)(w)$, and $\prob'\big((v,f)\big)\big((w,f)\big) = \prob(v)(w)$; and
 \item $\pri'(v)=\pri(v)$ for all $v \in \vertexSet$ and $\pri'\big((v,f)\big)=\pri(v)$ otherwise.
\end{itemize}
\end{definition}
Intuitively, the transformation changes the game so that, every time $R$ is entered, Player~$0$ has to fix her memoryless strategy in the game.
The fact that in the resulting game the strategy $f$ for Player~$0$ is fixed entering $R$ is due to the jump from the original vertex $v$ to $(v,f)$ whenever $v \in R$.
Once in $R$, either the part $v$ of $(v,f)$ is under the control of Player~$1$ or Player random, i.e. $v \notin \vertexSet_{0}$, so it behaves as in $\mg$, or the next state $w$ (or $(w,f)$ if $w \in R$) is the outcome of $f$, i.e. $w = f(v)$.

It is quite obvious that this transformation does not impact on the likelihood of winning.
In fact, Player~$0$ can simulate every memoryless strategy $f:\vertexSet_{0} \rightarrow \vertexSet$ by playing a strategy $f_R: \vertexSet_{0}' \rightarrow \vertexSet'$ that copies $f$ outside of $R$ (i.e.\ for each $v \in \vertexSet_{0} \setminus R$, $f_R(v) = f(v)$) and moves to the $f\restriction_R$ (i.e.\ $f$ with a preimage restricted to $R$) copy from states in $R$ (i.e.\ for each $v \in \vertexSet_{0} \cap R$, $f_R(v) = (v,f\restriction_R)$):
there is a one-to-one correspondence between playing in $\mg$ with strategy $f$ and playing in $\mg^R$ with strategy $f_R$ when starting in $\vertexSet$.

\begin{theorem}
\label{theo:PR}For all $v \in \vertexSet$, all $R \subseteq \vertexSet$, and all memoryless Player~$0$ strategies $f$, $\val_f^{\mg}(v) = \val_{f_R}^{\mg^R}\big((v,f\restriction_R)\big)$, $\val^{\mg}(v) = \sup\limits_{f\in \strats_0(\mg)}\val^{\mg^R}\big((v,f\restriction_R)\big)$, and $\val^{\mg}(v) = \val^{\mg^R}(v)$ hold.
\end{theorem}

\subsection{Correctness proof}
For a given game $\mg$, we call an $\epsilon \in (0,1)$ \emph{small} if it is at most $\frac{1}{5}$ of the smallest difference between all probabilities of winning that can occur on any strategy pair for any state in any game $\mg^R$ for any $R \subseteq \vertexSet$.
For every small $\epsilon$, we get the following corollary from Theorem \ref{theo:epsilon}.

\begin{corollary}[preservation of profitable and unprofitable switches]
\label{cor:preserve}
Let $n \geq |\mg|$, let $f$ be a Player~$0$ strategy for $\mg$, $f'$ the corresponding strategy for $\mg_{\epsilon,n}$, $\epsilon$ small, $v\in \vertexSet$, $w = f(v)$, and $(v,u) \in \edgeSet$.
Then $\val^{\mg}_f(u) > \val^{\mg}_f(w)$ implies $\val^{\mg_{\epsilon,n}}_{f'}(u) > \val^{\mg_{\epsilon,n}}_{f'}(w')$, and $\val^{\mg}_f(u) < \val^{\mg}_f(w)$ implies $\val^{\mg_{\epsilon,n}}_{f'}(u) < \val^{\mg_{\epsilon,n}}_{f'}(w')$.
\end{corollary}

It immediately follows that all combinations of profitable switches can be applied, and will lead to an improved strategy:
for small $\epsilon$, a profitable switch for $f_i$ from $f_i(v)=w$ to $f_{i+1}(v)=u$ implies $\val^{\mg}_{f_i}(u) \geq \val^{\mg}_{f_i}(w) + 5 \epsilon$ since by definition, we have that $\val^{\mg}_{f_i}(u) > \val^{\mg}_{f_i}(w)$ (as the switch is profitable); 
in particular, $\val^{\mg}_{f_i}(u) = \val^{\mg}_{f_i}(w) + \delta$ with $\delta \in \strictposreals$;
since $\epsilon \leq \frac{1}{5} \delta$, we have that $\val^{\mg}_{f_i}(u) \geq \val^{\mg}_{f_i}(w) + 5 \epsilon$.
The triangular inequalities provided by Theorem~\ref{theo:epsilon} imply that  $\val^{\mg_{\epsilon,n}}_{f_i'}(u') \geq \val^{\mg_{\epsilon,n}}_{f_i'}(w') + 3 \epsilon$, since $\big|\val^{\mg}_{f_i} - \val^{\mg_{\epsilon,n}}_{f_i'} \big| < \epsilon$.
Consequently, since under $f'_{i+1}$ we have that $\val^{\mg_{\epsilon,n}}_{f_{i+1}'}(v') = \val^{\mg_{\epsilon,n}}_{f_i'}(u')$, it follows that $\val^{\mg_{\epsilon,n}}_{f_{i+1}'}(v) \geq \val^{\mg_{\epsilon,n}}_{f_i'}(v) + 3 \epsilon$, and, using triangulation again, we get $\val^{\mg}_{f_{i+1}}(v) \geq \val^{\mg}_{f_i}(v) + \epsilon$. Thus, we have the following corollary:

\begin{corollary}
\label{cor:profitable}
Let $\mg$ be a given 2.5 player parity game, and $f_i$  be a  strategy with profitable switches ($\profit(\mg,f_i) {\neq} \emptyset$). Then, $\I_{f_i} \neq \emptyset$, and for all $f_{i+1} \in \I_{f_i}$, $\val^\mg_{f_{i+1}}{>}\val^\mg_{f_i}$.
\end{corollary}

We now turn to the case that there are no profitable switches for $f$ in the game $\mg$.
Corollary \ref{cor:preserve} shows that, for the corresponding strategy $f'$ in $\mg_{\epsilon,n}$, all profitable switches lie within the neutral edges for $f$ in $\mg$, provided $f$ has no profitable switches.

We expand the game by fixing the strategy of Player~$0$ for the vertices in $R \cap V_0$ for a region $R\subseteq \vertexSet$.
The region we are interested in is the winning region of Player $0$ in the neutral subgame $\mathcal P \cap \neutral(P,f)$.
The game is played as follows.

For every strategy $f_R \colon R \cap \vertexSet_0 \to \vertexSet$ such that $\big(r,f_R(r)\big) \in E$ holds for all $r \in R$, the game has a copy of the original game intersected with $R$, where the choices of Player~$0$ on the vertices in $R$ are fixed to the single choice defined by the respective strategy $f_R$.
We define $\|\mg\| = \max\setcond{|\mg^R|}{R \subseteq \vertexSet}$.

We consider the case where the almost sure winning region of Player~$0$ in the neutral subgame $\mg'=\mg \cap \neutral(\mg,f_i)$ is strictly larger than her winning region in $\mg_{f_i}$.
\begin{lemma}
\label{lem:asImprove}
Let $\mg$ be a given 2.5 player parity game,  and $f_i$ be a strategy such that the winning region $W_0'$ for Player $0$ in the neutral subgame $\mg'{=}\mg {\cap} \neutral(\mg,f_i)$ is strictly larger than her winning region $W_0$ in $\mg_{f_i}$. Then $\I_{f_i}' {\neq} \emptyset$ and, $\forall f_{i+1} {\in} \I_{f_i}'$, $\val^\mg_{f_{i+1}}{>}\val^\mg_{f_i}$.
\end{lemma}

\begin{myproof}
  The argument is an extension of the common argument for strategy improvement made for the modified reachability game.
  We first recall that the strategies in $\I_{f_i}'$ differ from $f_i$ only on the winning region $W_0'$ of Player $0$ in the neutral subgame $\mg'$.
  Assume that we apply the change \emph{once}: the first time $W_0'$ is entered, we play the new strategy, and after it is left, we play the old strategy.
  If the reaction of Player~$1$ is to stay in $W_0'$, Player~$0$ will win almost surely in $\mg$.
  If he leaves it, the value is improved due to the fact that Player $1$ has to take a disadvantageous edge to leave it.

Consider the game $\mg^{W_0'}$ and fix $f_{i+1} \in \I_{f_i}'$.  Using Theorem \ref{theo:PR}, this implies that, when first in a state $v \in W_0'$, Player $0$ moves to $(v,f_{i+1})$ for some $f_{i+1} \in \I_{f_i}'$, then the likelihood of winning is either improved or $1$ for any counter strategy of Player $1$.
  For all $v \in W_0' \setminus W_0$, this implies a strict improvement.
  For an $n \geq \|\mg\|$ and a small $\epsilon$, we can now follow the same arguments as for the Corollaries \ref{cor:preserve} and \ref{cor:profitable} on $\mg^{W_0'}$ to establish that $\val^{\mg^{W_0'}}_{(f_{i+1})_{W_0'}} > \val^{\mg^{W_0'}}_{(f_i)_{W_0'}}$ holds, where the inequality is obtained through the same steps:
  $\val^{\mg^{W_0'}}_{(f_i)_{W_0'}}\big((v,f_{i+1}|_{W_0})\big) > \val^{\mg^{W_0'}}_{(f_{i})_{W_0'}}(v) $ implies $\val^{\mg^{W_0'}}_{(f_i)_{W_0'}}\big((v,f_{i+1}|_{W_0})\big) \geq \val^{\mg^{W_0'}}_{(f_{i})_{W_0'}}(v) + 5 \epsilon$; this implies  $\val^{\mg^{W_0'}_{\epsilon,n}}_{(f_i)_{W_0'}}\big((v,f_{i+1}|_{W_0})'\big) \geq \val^{\mg^{W_0'}_{\epsilon,n}}_{(f_{i})_{W_0'}}(v) + 3 \epsilon$; and this implies $\val^{\mg^{W_0'}_{\epsilon,n}}_{(f_{i+1})_{W_0'}}(v) = \val^{\mg^{W_0'}_{\epsilon,n}}_{(f_{i+1})_{W_0'}}\big((v,f_{i+1}|_{W_0})'\big) \geq \val^{\mg^{W_0'}_{\epsilon,n}}_{(f_{i})_{W_0'}}(v) + 3 \epsilon$ and we finally get
  $\val^{\mg^{W_0'}}_{(f_{i+1})_{W_0'}}(v) =\val^{\mg^{W_0'}}_{(f_{i+1})_{W_0'}}\big((v,f_{i+1}|_{W_0})\big)  > \val^{\mg^{W_0'}}_{(f_{i})_{W_0'}}(v)$.

  With Theorem \ref{theo:PR}, we obtain that $\val^{\mg}_{f_{i+1}} > \val^{\mg}_{f_i}$ holds.
\end{myproof}

Let us finally consider the case where there are no profitable switches for Player $0$ in $\mg_{f_i}$ and her winning region on the neutral subgame $\mg \cap \neutral(\mg,f_i)$ coincides with her winning region in $\mg_{f_i}$.

\begin{lemma}
\label{lem:noImprovement}
Let $\mg$ be an MPG and $f_i$ be a strategy such that the set of profitable switches is empty and the neutral subgame $\mg \cap \neutral(\mg,f_i)$ has the same winning region for Player $0$ as her winning region in $\mg_{f_i}$ ($\I_{f_i} = \I_{f_i}' = \emptyset$). Then, every individual profitable switch in the reachability game $\mg_{\epsilon,n}$ from $f_i$ to $f_{i+1}$ implies $\val^\mg_{f_{i+1}} = \val^\mg_{f_i}$ and $\neutral(\mg,f_{i+1}) = \neutral(\mg,f_i)$.
\end{lemma}

\begin{myproof}
When there are no profitable switches in the parity game $\mg$ for $f_i$, then all profitable switches in the reachability game $\mg_{\epsilon,n}$ for $f_i$ (if any) must be within the set of neutral edges $\neutral(\mg,f_i)$ in the parity game $\mg$.
We apply one of these profitable switches at a time. 
By our definitions, this profitable switch is neutral in the 2.5 player parity game.

Taking this profitable (in the reachability game $\mg_{\epsilon,n}$ for a small $\epsilon$ and some $n \geq \|\mg\|$) switch will improve the likelihood of winning for Player~$0$ in the reachability game.
By our definition of $\epsilon$, this implies that the likelihood of winning cannot be decreased on any position in the parity game.

To see that the quality of the resulting strategy cannot be higher for Player~$0$ in the 2.5 player parity game, recall that Player~$1$ can simply follow his optimal strategy on the neutral subgame.
The likelihood of winning for Player $0$ is the likelihood of reaching her winning region, and this winning region has not changed.
Moreover, consider the evaluation of the likelihood of reaching this winning region:
since by fixing the strategy for Player~$1$ the resulting game is an MDP, such an evaluation can be obtained by solving a linear programming problem
\ifarxiv
(cf. the arXiv version for more details).
\else
(sketched in Appendix~\ref{apx:reach-lp}).
\fi
The old minimal non-negative solution to the resulting linear programming problem is a solution to the new linear programming problem, as it satisfies all constraints.

Putting these arguments together, likelihood of winning in the parity game is not altered in any vertex by this change.
Hence, the set of neutral edges is not altered.
\end{myproof}

This lemma implies that \emph{none} of the subsequently applied improvement steps applied on the 2.5 player reachability game has any effect on the quality of the resulting strategy on the 2.5 player parity game.
Together, the above lemmas and corollaries therefore provide the correctness argument.

\begin{theorem}
  The algorithm is correct.
\end{theorem}

\begin{myproof}
  Lemma \ref{lem:noImprovement} shows that, when $\I_{f_i}$ and $\I_{f_i}'$ are empty (i.e.\ when the algorithm terminates), then the updates in the related 2.5 player reachability game will henceforth (and thus until termination) not change the valuation for the 2.5 player parity game. 
  With Theorems \ref{theo:SI4reach} and \ref{theo:epsilon} and our selection of small $\epsilon$, it follows that $f_i$ is an optimal strategy.
  The earlier lemmas and corollaries in this subsection show that every strategy $f_{i+1} \in \I_{f_i} \cup \I_{f_i}'$ satisfies $\val^\mg_{f_{i+1}} > \val^\mg_{f_{i}}$.
  Thus, the algorithm produces strategies with strictly increasing quality in each step until it terminates.
As the game is finite, then also the set of strategies is finite, thus the algorithm will terminate after finitely many improvement steps with an optimal strategy.
\end{myproof}

As usual with strategy improvement algorithms, we cannot provide good bounds on the number of iterations.
As reachability games are a special case of 2.5 player games, all selection rules considered by Friedmann \cite{Friedmann/09/lower,FHZ/11b/lower} will have exponential lower bounds.

\section{Implementation and experimental results}
\label{sec:implementation}

We have written a prototypical implementation for the approach of this paper.
Our tool supports the input language of the probabilistic model checker \prismgames~\cite{ChenFKPS13}, an extension of \prism~\cite{KwiatkowskaNP11} to stochastic Markov games.
As case study, we consider a battlefield consisting of $n \times n$ square tiles, surrounded by a solid wall.
On the battlefield there are two robots, $R_{0}$ and $R_{1}$, and four marked zones $\mathrm{zone}_{1}, \ldots, \mathrm{zone}_4$ at the corners, each of size $3 \times 3$.
Each tile can be occupied by at most one robot at a time.
The robots act in strict alternation.
When it is the turn of a robot, this robot can move as follows:
decide a direction and move one field forward;
decide a direction and attempt to move two fields forward.
In the latter case, the robot moves two fields forward with a probability of 50\%, but only one field forward with a probability of 50\%.
If the robot would run into a wall or into the other robot, it stops at the field before the obstacle.
Robot $R_{1}$ can also shoot $R_{0}$ instead of moving, which is destroyed with probability $p_{\mathit{destr}}^{d}$ where $p_{\mathit{destr}}$ is the probability of destroying the robot and $d$ is the Euclidean distance between the two robots.
Once destroyed, $R_{0}$ cannot move any more.
We assume that we are in control of $R_{0}$ but cannot control the behaviour of $R_{1}$.
Our goal is to maximise, under any possible behaviour of $R_{1}$, the probability of fulfilling a certain objective depending on the zones, such as repeatedly visiting all zones infinitely often, visiting the zones in a specific order, performing such visits without entering other zones in the meanwhile, and so on.
As an example, we can specify that the robot eventually reaches each zone by means of the
probabilistic LTL (PLTL) formula $\llangle R_{0} \rrangle \ltlPmax [\bigwedge_{i=1,\dots,4}\ltlF \, \mathrm{zone}_{i}]$ requiring to maximise the probability of satisfying $\bigwedge_{i=1,\dots,4}\ltlF \, \mathrm{zone}_{i}$ by controlling $R_{0}$ only.

\begin{table*}[t]
	\caption{Robots analysis: different reachability properties}
	\label{tab:robots}	
	\centering
	\setlength{\tabcolsep}{4pt}
	\resizebox{\textwidth}{!}{
	\begin{tabular}{lrc|rr|rr|rr|rr|rr|rr}
		\toprule
		\multicolumn{1}{c}{\multirow{2}{*}{property}} & \multirow{2}{*}{$n$} & \multirow{2}{*}{$b$} & \multicolumn{2}{c|}{MPG} & \multicolumn{2}{c|}{$p_{\mathit{destr}} = 0.1$} & \multicolumn{2}{c|}{$p_{\mathit{destr}} = 0.3$} & \multicolumn{2}{c|}{$p_{\mathit{destr}} = 0.5$} & \multicolumn{2}{c|}{$p_{\mathit{destr}} = 0.7$} & \multicolumn{2}{c}{$p_{\mathit{destr}} = 0.9$} \\
		& & & \multicolumn{1}{c}{vertices} & colours & \multicolumn{1}{c}{$p_{\max}$} & $t_{\mathit{sol}}$ & \multicolumn{1}{c}{$p_{\max}$} & $t_{\mathit{sol}}$ & \multicolumn{1}{c}{$p_{\max}$} & $t_{\mathit{sol}}$ & \multicolumn{1}{c}{$p_{\max}$} & $t_{\mathit{sol}}$ & \multicolumn{1}{c}{$p_{\max}$} & $t_{\mathit{sol}}$\\
		\midrule
		\multirow{5}{*}{
		\begin{tabular}{l}
		Reachability\\$\llangle R_{0}\rrangle\ltlPmax$\\
		$[ \hphantom{{}\wedge{}} \ltlF  \mathrm{zone}_{1} \wedge \ltlF \mathrm{zone}_2 \hphantom{]} $\\
		$\hphantom{[} \wedge \ltlF \mathrm{zone}_3  \wedge \ltlF \mathrm{zone}_4 ]$
		\end{tabular}
		}
		&  7 & 1 &    663\,409 & 2 & 0.9614711 &  33 & 0.8178044 &  22 & 0.6247858 &  22 & 0.3961410 &  21 & 0.1384328 &  23 \\
		&  7 & 2 & 1\,090\,537 & 2 & 0.9244309 &  56 & 0.6742610 &  66 & 0.4017138 &  57 & 0.1708971 &  58 & 0.0230085 &  52 \\
		&  7 & 3 & 1\,517\,665 & 2 & 0.8926820 &  89 & 0.5793073 &  87 & 0.2995397 &  77 & 0.0953904 &  86 & 0.0060025 &  68 \\
		&  7 & 4 & 1\,944\,793 & 2 & 0.8667039 & 112 & 0.5385632 & 109 & 0.2409219 &  96 & 0.0649772 & 107 & 0.0026513 &  85 \\
		&  7 & 5 & 2\,371\,921 & 2 & 0.8571299 & 147 & 0.5062357 & 144 & 0.2167625 & 127 & 0.0506530 & 140 & 0.0019157 & 112 \\
		\midrule
		\multirow{5}{*}{
		\begin{tabular}{l}
		Ordered\\Reachability\\$\llangle R_{0}\rrangle\ltlPmax$\\
		$[ \ltlF (\mathrm{zone}_{1} \wedge \ltlF \, \mathrm{zone}_2) ]$
		\end{tabular}
		}
		&  8 & 1 &    528\,168 & 2 & 0.9613511 &  23 & 0.8176058 &  19 & 0.6246643 &  21 & 0.3962011 &  20 & 0.1384974 &  19 \\
		&  8 & 2 &    868\,986 & 2 & 0.9243652 &  35 & 0.6999023 &  44 & 0.4522051 &  35 & 0.2083732 &  42 & 0.0320509 &  40 \\
		&  8 & 3 & 1\,209\,804 & 2 & 0.9091132 &  62 & 0.6538475 &  71 & 0.3643938 &  56 & 0.1352710 &  60 & 0.0131408 &  58 \\
		&  8 & 4 & 1\,550\,622 & 2 & 0.9013742 &  91 & 0.6200998 &  91 & 0.3316778 &  72 & 0.1168758 &  74 & 0.0097312 &  71 \\
		&  8 & 5 & 1\,891\,440 & 2 & 0.8977303 & 113 & 0.6031945 & 108 & 0.3207408 &  90 & 0.1138603 &  88 & 0.0093679 &  83 \\
		\midrule
		\multirow{5}{*}{
		\begin{tabular}{l}
		Reach-Avoid\\$\llangle R_{0}\rrangle\ltlPmax$\\
		$[ \hphantom{{} \wedge {}} \neg \mathrm{zone}_{1} \ltlU \mathrm{zone}_2 $\\
		$\hphantom{[} \wedge \neg \mathrm{zone}_4 \ltlU \mathrm{zone}_2 $\\
		$\hphantom{[} \wedge \ltlF \mathrm{zone}_3 ]$
		\end{tabular}
		}
		&  9 & 1 &    833\,245 & 4 & 0.9447793 &  46 & 0.8005413 &  31 & 0.6125397 &  35 & 0.3914531 &  25 & 0.1372075 &  24 \\
		&  9 & 2 & 1\,370\,827 & 4 & 0.9095579 &  81 & 0.6824329 &  52 & 0.4411181 &  61 & 0.2089446 &  49 & 0.0302023 &  45 \\
		&  9 & 3 & 1\,908\,409 & 4 & 0.8972146 & 108 & 0.6375883 &  68 & 0.3792906 &  84 & 0.1444959 &  71 & 0.0106721 &  66 \\
		&  9 & 4 & 2\,445\,991 & 4 & 0.8936231 & 148 & 0.6221536 &  93 & 0.3478172 & 117 & 0.1158094 & 103 & 0.0051508 &  89 \\
		&  9 & 5 & 2\,983\,573 & 4 & 0.8918034 & 172 & 0.6162166 & 109 & 0.3366050 & 136 & 0.1010400 & 120 & 0.0035468 & 105 \\
		\midrule
		\multirow{5}{*}{
		\begin{tabular}{l}
		Reachability\\$\llangle R_{0}\rrangle\ltlPmax$\\
		$[ \hphantom{{}\wedge{}} \ltlF  \mathrm{zone}_{1} \wedge \ltlF \mathrm{zone}_2 \hphantom{]} $\\
		$\hphantom{[} \wedge \ltlF \mathrm{zone}_3  \wedge \ltlF \mathrm{zone}_4 ]$
		\end{tabular}
		}
		& 10 & 1 &  3\,307\,249 & 2 & 0.9614711 & 186 & 0.8178044 & 141 & 0.6247858 & 142 & 0.3961410 &  142 & 0.1384328 &  141 \\
		& 10 & 2 &  5\,440\,429 & 2 & 0.9244267 & 296 & 0.6755372 & 414 & 0.4017718 & 374 & 0.1665626 &  732 & 0.0207851 &  615 \\
		& 10 & 3 &  7\,573\,609 & 2 & 0.8931881 & 570 & 0.5742127 & 572 & 0.2864117 & 509 & 0.0847474 & 1019 & 0.0043153 &  861 \\
		& 10 & 4 &  9\,706\,789 & 2 & 0.8676441 & 530 & 0.5239018 & 794 & 0.2248369 & 735 & 0.0479367 & 1396 & 0.0009959 & 1610 \\
		& 10 & 5 & 11\,839\,969 & 2 & 0.8503684 & 968 & 0.4885654 & 980 & 0.1866995 & 971 & 0.0305890 & 1708 & \multicolumn{2}{c}{---TO---} \\
		\bottomrule
	\end{tabular}
	}
\end{table*}

The machine we used for the experiments is a 3.6 GHz Intel Core i7-4790 with 16GB 1600 MHz DDR3 RAM of which 12GB assigned to the tool;
the timeout has been set to 30 minutes.
We have applied our tool on a number of properties that require the robot $R_{0}$ to visit the different zones in a certain order.
In Table~\ref{tab:robots} we report the performance measurements for these properties.
Column ``property'' shows the PLTL formula we consider, column ``$n$'' the width of the battlefield instance, and column ``$b$'' the number of bullets $R_{1}$ can shoot.
For the ``MPG'' part, we present the number of ``vertices'' of the resulting MPG and the number of ``colours''.
In the remaining columns, for each value of ``$p_{\mathit{destr}}$'', we report the achieved maximum probability ``$p_{\max}$'' and the time ``$t_{\mathit{sol}}$'' in seconds needed to solve the game.
Note that we cannot compare to \prismgames because it does not support general PLTL formulas, and we are not aware of other tools to compare with.

As we can see, the algorithm performs quite well on MPGs with few million states.
It is worth mentioning that a large share of the time spent is due to the evaluation of the 1.5 player parity games in the construction of the profitable switches.
For instance, such an evaluation required 137 seconds out of 172 for the case $n = 9$, $b = 5$, and $p_{\mathit{destr}} = 0.1$.
Since a large part of these 1.5 player games are similar, we are investigating how to avoid the repeated evaluation of similar parts to reduce the running time.
Generally, all improvements in the quantitative solution of 1.5 player parity games and the qualitative solution of 2.5 player parity games will reduce the running time of our algorithm.

\section{Discussion}

We have combined a recursive algorithm for the quantitative solution of 2.5 player parity games with a strategy improvement algorithm, which lifts these results to the qualitative solution of 2.5 player parity games.
This shift is motivated by the significant acceleration in the qualitative solution of 2.5 player parity games: while \cite{DBLP:conf/cav/ChatterjeeHJR10} scaled to a few thousand vertices, \cite{HahnSTZ16} scales to tens of millions of states.
This changes the playing field and makes qualitative synthesis a realistic target.
It also raises the question if this technique can be incorporated smoothly into a quantitative solver.

Previous approaches \cite{DBLP:conf/stacs/ChatterjeeH06,ChatterjeeAH06} have focused on developing a progress measure that allows for joining the two objective.
This has been achieved in studying strategy improvement techniques that give preference to the likelihood of winning, and overcome stalling by performing strategy improvement on the larger qualitative game from \cite{chatterjee-04-paper} on the value classes.

This approach was reasonable at the time, where the updates benefited from memorising the recently successful strategies on the qualitative game.
Moreover, focussing on value classes keeps the part of the qualitative game under consideration small, which is a reasonable approach when the cost of qualitative strategy improvement is considered significant.
Building on a fast solver for the qualitative analysis, we can afford to progress in larger steps.

The main advancement, however, is as simple as it is effective.
We use strategy improvement where it has a simple direct meaning (the likelihood to win), and we do not use it where the progress measure is indirect (progress measure within a value class).
This has allowed us to transfer the recent performance gains from qualitative solutions of 2.5 player parity games \cite{HahnSTZ16} to their quantitative solution.

The difference in performance also explains the difference in the approach regarding complexity.
Just as the deterministic subexponential complexity of solving 2.5 player games qualitatively is not very relevant in \cite{HahnSTZ16} (as this approach would be very slow in practice), the expected subexponential complexity in \cite{DBLP:conf/stacs/ChatterjeeH06} is bought by exploiting a random facet method, which implies that only one edge is updated in every step.
From a theoretical angle, these complexity considerations are interesting.
From a practical angle, however, strategy improvement algorithms that use multiple switches in every step are usually faster and therefore preferable.

\bibliographystyle{IEEEtran}
\bibliography{bib}

\ifarxiv
\newpage
\appendix
\section{Algorithm}
\label{apx:algorithm}

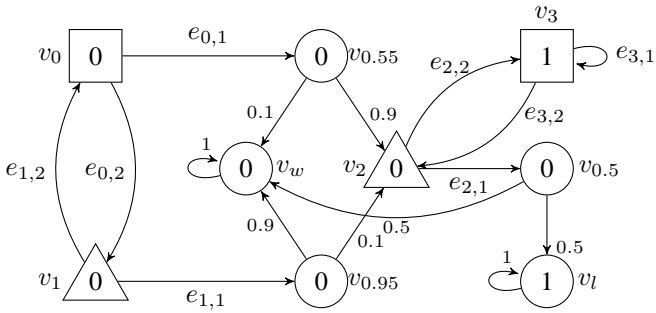
\begin{figure}
  \centering
  \begin{tikzpicture}[->,>=stealth',auto]
    \node[p0node] (p0) at (0,0) {0};
    \node at ($(p0)+(-0.6,0)$) {$v_{0}$};
    \node[p1node] (p1) at (0,-3) {0};
    \node at ($(p1)+(-0.6,0)$) {$v_{1}$};
    \node[prnode] (prob0) at (3,0) {0};
    \node at ($(prob0)+(0.7,0)$) {$v_{0.55}$};
    \node[prnode] (prob1) at (3,-3) {0};
    \node at ($(prob1)+(0.7,0)$) {$v_{0.95}$};
    \node[prnode] (ptarget) at (2,-1.5) {0};
    \node at ($(ptarget)+(0.6,0)$) {$v_{w}$};
    \node[p1node] (p2) at (4,-1.5) {0};
    \node at ($(p2)+(-0.55,0)$) {$v_{2}$};
    \node[p0node] (p3) at (6,0) {1};
    \node at ($(p3)+(0,0.55)$) {$v_{3}$};
    \node[prnode] (prob2) at (6,-1.5) {0};
    \node at ($(prob2)+(0.7,0)$) {$v_{0.5}$};
    \node[prnode] (prob3) at (6,-3) {1};
    \node at ($(prob3)+(0.55,0)$) {$v_{l}$};

    \draw (p0) to[bend left] node[left] {$e_{0,2}$} (p1);
    \draw (p1) to[bend left] node {$e_{1,2}$} (p0);
    \draw (p0) to node {$e_{0,1}$} (prob0);
    \draw (p1) to node[below] {$e_{1,1}$} (prob1);
    \draw (prob0) to node[left] {\scriptsize$0.1$} (ptarget);
    \draw (prob0) to node[right] {\scriptsize$0.9$} (p2);
    \draw (prob1) to node[left] {\scriptsize$0.9$} (ptarget);
    \draw (prob1) to node[right,near start] {\scriptsize$0.1$} (p2);
    \draw (ptarget) edge[loop left] node[above,very near end] {\scriptsize$1$} (ptarget);
    \draw (p2) to[bend left] node[above] {$e_{2,2}$} (p3);
    \draw (p2) to node[below] {$e_{2,1}$} (prob2);
    \draw (p3) to[bend left] node[right,near start] {$e_{3,2}$} (p2);
    \draw (p3) edge[loop right] node[right] {$e_{3,1}$} (p3);
    \draw (prob2) to node[right,very near end] {\scriptsize$0.5$} (prob3);
    \draw (prob2) to[bend left=29,inner sep=1pt] node[below] {\scriptsize$0.5$} (ptarget);
    \draw (prob3) edge[loop left] node[above,very near end] {\scriptsize$1$} (prob3);
  \end{tikzpicture}
  \caption{Extended probabilistic parity game $\mg_{x}$.}
  \label{fig:example-extended}
\end{figure}

In this section we provide the details of the algorithm presented in the main part of the paper.
It is an implementation of the algorithm we have described in Section \ref{sec:algorithm} and contains our design decisions. They are not relevant for correctness. We consider the extended game in Figure~\ref{fig:example-extended} for our running example.

We start with an initialisation, where we solve 2.5 player parity games qualitatively, and may require the qualitative solution of subgames.
For this initialisation, we first define an extended qualitative solution of 2.5 player game as a memoryless strategy for a player, which guarantees that s/he wins almost surely on his or her almost sure winning region \emph{and} only loses almost surely on the almost sure winning region of his or her opponent (cf.~Section~\ref{ssec:initialisation}).

We initialise our strategy improvement algorithm with an extended qualitative solution.
To obtain such a solution, we can essentially use the algorithms known from ordinary qualitative solutions, cf.\ Definition~\ref{def:aux_algos}.

The \emph{normal improvement step} is an instance of standard strategy improvement algorithms.
We evaluate the likelihood of winning for Player~$0$ for her current strategy, %
by computing the value optimal for Player~$1$ against this strategy.
If we can obtain an improvement by changing a decision in a Player~$0$ state, we do so.

The correctness proof in the main part of the paper uses the related 2.5 player reachability game as a comparison point in the correctness argument.
It shows that, for sufficiently small $\epsilon$, each improvement selected by the algorithm is also an improvement in the related reachability game, while the stopping condition guarantees that further improvements in the reachability games do not translate to further improvements in the Markov parity game.

Consequently, our technique not only avoids using the tiny $\epsilon$, it also avoids the slow progression through updates that are stale (lead to no improvement) on the parity game while leading to an improvement on the reachability game resulting from the translation.

Our main algorithm is given as \Call{Main}{$\functionDot$} in Algorithm~\ref{alg:main}.
It makes use of the auxiliary algorithms from Definition~\ref{def:aux_algos}.

\begin{definition}
  \label{def:aux_algos}
  For an MPG $\mg = (\vertexSet_{0}, \vertexSet_{1}, \vertexSet_{r}, \edgeSet, \pri)$ we denote by $\Call{QualiSolve}{\mg} = (\winreg,f)$ a method which computes the winning regions $\winreg$ of Player~$0$ and a Player~$0$ strategy $f$ winning in $\winreg$ and arbitrary defined elsewhere.

  Further, for $A \subseteq \vertexSet$, we let $\Call{Reach}{\mg, A} = \val$ denote the result of computing mutually optimal reachability probabilities, that is
  $\val = \val^{\mg'}$ where $\mg'$ is the reachability game $\mg' = (\vertexSet_{0}, \vertexSet_{1}, \vertexSet_{r}, \edgeSet, A)$.

  For an MPG $\mg = (\vertexSet_{0}, \emptyset, \vertexSet_{r}, \edgeSet, \pri)$, whose arena is an MDP, we let $\Call{EvaluateMDP}{\mg} = \val$ denote the value of the MDP, that is
$
  \val = \val^{\mg} .
$
\end{definition}
\Call{QualiSolve}{$\mg$} can be effectively implemented by \cite{HahnSTZ16} without having to construct intermediate 2 player games.

Note that efficient procedures for evaluating parity MDPs exist~\cite{Courcoubetis+Yannakakis/95/Markov}.
Having to control only a single player, such algorithms merely need to determine the almost sure winning region of this single player (which is simple) and compute the maximal probability to reach this region (cf.\ Appendix \ref{apx:reach-lp}).
(This does not hold for general 2.5 player parity games.)

In our setting, we obtain a parity MDP by fixing the strategy for Player~$0$.
We therefore have to compute the \emph{minimal} values for reachability.
We can, however, transform this parity game by adding $1$ to the parity labels, so to complement the winning condition, computing the maximal winning values for such a complement, and finally returning $1$ minus the value computed for each node.

\begin{algorithm*}
\caption{Quantitative parity game solving algorithm}
\label{alg:main}
\begin{algorithmic}[1]
  \Function {Main}{$\mg$}%
  \begin{tikzpicture}[->,>=stealth',auto,scale=0.555, every node/.style={transform shape}]
	\path[use as bounding box] (-16.75,1) rectangle (-16.75,1);
	\node at (3,1) {\LARGE$\Call{Initialise}{\mg_{x}}$, at line~\ref{algline:init-empty-return-begin}};
    \node[p0node] (p0) at (0,0) {0};
    \node[p1node] (p1) at (0,-3) {0};
    \node[prnode] (prob0) at (3,0) {0};
    \node[prnode] (prob1) at (3,-3) {0};
    \node[prnode] (ptarget) at (2,-1.5) {0};
    \node[p1node] (p2) at (4,-1.5) {0};
    \node[p0node] (p3) at (6,0) {1};
    \node[prnode] (prob2) at (6,-1.5) {0};
    \node[prnode] (prob3) at (6,-3) {1};

    \draw[dashed] ($(ptarget)-(1.25,0.5)$) rectangle ($(ptarget)+(0.5,0.5)$);
    \node at ($(ptarget)-(1,0.25)$) {$\winreg$};

    \draw (p0) to[bend left] (p1);
    \draw[dashed] (p1) to[bend left] (p0);
    \draw[dotted] (p0) to (prob0);
    \draw[dashed] (p1) to (prob1);
    \draw[dashdotted] (prob0) to (ptarget);
    \draw[dashdotted] (prob0) to (p2);
    \draw[dashdotted] (prob1) to (ptarget);
    \draw[dashdotted] (prob1) to (p2);
    \draw[dashdotted] (ptarget) edge[loop left] (ptarget);
    \draw[dashed] (p2) to[bend left] (p3);
    \draw[dashed] (p2) to (prob2);
    \draw[dotted] (p3) to[bend left] (p2);
    \draw (p3) edge[loop right] (p3);
    \draw[dashdotted] (prob2) to (prob3);
    \draw[dashdotted] (prob2) to[bend left] (ptarget);
    \draw[dashdotted] (prob3) edge[loop left] (prob3);
  \end{tikzpicture}%
  \begin{tikzpicture}[->,>=stealth',auto,scale=0.555, every node/.style={transform shape}]
	\path[use as bounding box] (-16.75,7) rectangle (-16.75,7);
	\node at (3,1) {\LARGE$\Call{Initialise}{\mg_{x}}$, at line~\ref{algline:init-improve-end}};
    \node[p0node] (p0) at (0,0) {0};
    \node[p1node] (p1) at (0,-3) {0};
    \node[prnode] (prob0) at (3,0) {0};
    \node[prnode] (prob1) at (3,-3) {0};
    \node[prnode] (ptarget) at (2,-1.5) {0};
    \node[p1node] (p2) at (4,-1.5) {0};
    \node[p0node] (p3) at (6,0) {1};
    \node[prnode] (prob2) at (6,-1.5) {0};
    \node[prnode] (prob3) at (6,-3) {1};

    \draw[dashed] ($(ptarget)-(1.25,0.5)$) rectangle ($(ptarget)+(0.5,0.5)$);
    \node at ($(ptarget)-(1,0.25)$) {$\winreg$};

    \draw[dotted] (p0) to[bend left] (p1);
    \draw[dashed] (p1) to[bend left] (p0);
    \draw (p0) to (prob0);
    \draw[dashed] (p1) to (prob1);
    \draw[dashdotted] (prob0) to (ptarget);
    \draw[dashdotted] (prob0) to (p2);
    \draw[dashdotted] (prob1) to (ptarget);
    \draw[dashdotted] (prob1) to (p2);
    \draw[dashdotted] (ptarget) edge[loop left] (ptarget);
    \draw[dashed] (p2) to[bend left] (p3);
    \draw[dashed] (p2) to (prob2);
    \draw[dotted] (p3) to[bend left] (p2);
    \draw (p3) edge[loop right] (p3);
    \draw[dashdotted] (prob2) to (prob3);
    \draw[dashdotted] (prob2) to[bend left] (ptarget);
    \draw[dashdotted] (prob3) edge[loop left] (prob3);
  \end{tikzpicture}%
  \begin{tikzpicture}[->,>=stealth',auto,scale=0.555, every node/.style={transform shape}]
	\path[use as bounding box] (-16.75,13) rectangle (-16.75,13);
	\node at (3,1) {\LARGE$\Call{Initialise}{\mg_{x} \cap \setnocond{v_{l},v_{2},v_{3}}}$, at line~\ref{algline:init-empty-return-begin}};
    \node[p0node] (p0) at (0,0) {0};
    \node[p1node] (p2) at (4,-1.5) {0};
    \node[p0node] (p3) at (6,0) {1};

    \draw[dashed] ($(p2)-(0.75,0.5)$) rectangle ($(p3)+(1,0.5)$);
    \node at (4,0) {$\winreg$};

    \draw[dashed] (p2) to[bend left] (p3);
    \draw[dotted] (p3) to[bend left] (p2);
    \draw (p3) edge[loop right] (p3);
  \end{tikzpicture}%
  \begin{tikzpicture}[->,>=stealth',auto,scale=0.555, every node/.style={transform shape}]
	\path[use as bounding box] (-16.75,17.5) rectangle (-16.75,17.5);
	\node at (3,1) {\LARGE$\Call{Initialise}{\mg_{x} \cap \setnocond{v_{l},v_{2},v_{3}}}$, at line~\ref{algline:init-improve-end}};
    \node[p0node] (p0) at (0,0) {0};
    \node[p1node] (p2) at (4,-1.5) {0};
    \node[p0node] (p3) at (6,0) {1};

    \draw[dashed] ($(p2)-(0.75,0.5)$) rectangle ($(p3)+(1,0.5)$);
    \node at (4,0) {$\winreg$};

    \draw[dashed] (p2) to[bend left] (p3);
    \draw (p3) to[bend left] (p2);
    \draw[dotted] (p3) edge[loop right] (p3);
  \end{tikzpicture}%
  \begin{tikzpicture}[->,>=stealth',auto,scale=0.555, every node/.style={transform shape}]
	\path[use as bounding box] (-16.75,22) rectangle (-16.75,22);
	\node at (3,1) {\LARGE$\Call{Improve}{\mg_{x},f}$, at line~\ref{algline:improve-head}};
    \node[p0node] (p0) at (0,0) {0};
    \node[p1node] (p1) at (0,-3) {0};
    \node[prnode] (prob0) at (3,0) {0};
    \node[prnode] (prob1) at (3,-3) {0};
    \node[prnode] (ptarget) at (2,-1.5) {0};
    \node[p1node] (p2) at (4,-1.5) {0};
    \node[p0node] (p3) at (6,0) {1};
    \node[prnode] (prob2) at (6,-1.5) {0};
    \node[prnode] (prob3) at (6,-3) {1};

    \draw[dotted] (p0) to[bend left] (p1);
    \draw[dashed] (p1) to[bend left] (p0);
    \draw (p0) to (prob0);
    \draw[dashed] (p1) to (prob1);
    \draw[dashdotted] (prob0) to (ptarget);
    \draw[dashdotted] (prob0) to (p2);
    \draw[dashdotted] (prob1) to (ptarget);
    \draw[dashdotted] (prob1) to (p2);
    \draw[dashdotted] (ptarget) edge[loop left] (ptarget);
    \draw[dashed] (p2) to[bend left] (p3);
    \draw[dashed] (p2) to (prob2);
    \draw (p3) to[bend left] (p2);
    \draw[dotted] (p3) edge[loop right] (p3);
    \draw[dashdotted] (prob2) to (prob3);
    \draw[dashdotted] (prob2) to[bend left] (ptarget);
    \draw[dashdotted] (prob3) edge[loop left] (prob3);
  \end{tikzpicture}%
  \begin{tikzpicture}[->,>=stealth',auto,scale=0.555, every node/.style={transform shape}]
	\path[use as bounding box] (-16.75,28) rectangle (-16.75,28);
	\node at (3,1) {\LARGE$\Call{Improve}{\mg_{x},f}$, at line~\ref{algline:evaluate-update-end}};
    \node[p0node] (p0) at (0,0) {0};
    \node[p1node] (p1) at (0,-3) {0};
    \node[prnode] (prob0) at (3,0) {0};
    \node[prnode] (prob1) at (3,-3) {0};
    \node[prnode] (ptarget) at (2,-1.5) {0};
    \node[p1node] (p2) at (4,-1.5) {0};
    \node[p0node] (p3) at (6,0) {1};
    \node[prnode] (prob2) at (6,-1.5) {0};
    \node[prnode] (prob3) at (6,-3) {1};

    \draw (p0) to[bend left] (p1);
    \draw[dashed] (p1) to[bend left] (p0);
    \draw[dotted] (p0) to (prob0);
    \draw[dashed] (p1) to (prob1);
    \draw[dashdotted] (prob0) to (ptarget);
    \draw[dashdotted] (prob0) to (p2);
    \draw[dashdotted] (prob1) to (ptarget);
    \draw[dashdotted] (prob1) to (p2);
    \draw[dashdotted] (ptarget) edge[loop left] (ptarget);
    \draw[dashed] (p2) to[bend left] (p3);
    \draw[dashed] (p2) to (prob2);
    \draw (p3) to[bend left] (p2);
    \draw[dotted] (p3) edge[loop right] (p3);
    \draw[dashdotted] (prob2) to (prob3);
    \draw[dashdotted] (prob2) to[bend left] (ptarget);
    \draw[dashdotted] (prob3) edge[loop left] (prob3);
  \end{tikzpicture}%
  \State $f \gets \Call{Initialise}{\mg}$\label{algline:main-init}
  \State \Return \Call{Improve}{$\mg, f$}\label{algline:main-improve}
  \EndFunction
  \Statex
  \Function {Initialise}{$\mg$}
  \Statex where $\mg = (\vertexSet_{0}, \vertexSet_{1}, \vertexSet_{r}, \edgeSet, \prob, \pri)$ and $\vertexSet = \vertexSet_{0} \cup \vertexSet_{1} \cup \vertexSet_{r}$
  \State $(\winreg, f) \gets \Call{QualiSolve}{\mg}$\label{algline:init-almost}
  \If {$\winreg = \emptyset$}\label{algline:init-empty-return-begin}
    \State \Return f\label{algline:init-empty-return-mid}
  \EndIf\label{algline:init-empty-return-end}
  \State $(\val, g) \gets \Call{Reach}{\mg, \winreg}$\label{algline:init-reach}
  \For {$v \in \vertexSet_{0}\setminus \winreg$}\label{algline:init-improve-begin}
  \State $f(v) \gets g(v)$
  \EndFor\label{algline:init-improve-end}
  \State $\vertexSet' \gets \setcond{v \in \vertexSet}{\val(v) = 0}$\label{algline:init-recurse-begin}
  \State $g \gets \Call{Initialise}{\mg \cap \vertexSet'}$\label{algline:init-call-recurse}
  \For {$v \in \vertexSet_{0} \cap \vertexSet'$}
  \State $f(v) \gets g(v)$
  \EndFor\label{algline:init-recurse-end}
  \State \Return $f$
  \EndFunction
  \Statex
  \Function{Improve}{$\mg, f$}
  \Statex where $\mg = (\vertexSet_{0}, \vertexSet_{1}, \vertexSet_{r}, \edgeSet, \prob, \pri)$,
  \Statex \ \ $\vertexSet = \vertexSet_{0} \cup \vertexSet_{1} \cup \vertexSet_{r}$,
  \Statex  \ \
  $\edgeSet_{0} = E \cap \vertexSet_{0} {\times} \vertexSet$,
  \Statex \ \ $f$ is a 
  strategy
  \label{algline:improve-head}
  \Repeat
  \State $g \gets f$
  \State $\mg' \gets \mg_{f}$\label{algline:construct-mdp}
  \State $\val \gets \Call{EvaluateMDP}{\mg'}$\label{algline:evaluate-mdp}
  \For {$(v,v') \in \edgeSet_{0}$ with $\val(v') > \val(v)$}\label{algline:evaluate-profitable-switches-begin}
  \State $f(v) \gets v'$
  \EndFor\label{algline:evaluate-profitable-switches-end}
  \If {$f = g$}\label{algline:evaluate-no-profitable-switches-begin}
  \State $\mg' \gets \mg \cap \neutral(\mg, \val)$\label{algline:evaluate-neutral}
  \State $(\winreg, h) \gets \Call{QualiSolve}{\mg'}$\label{algline:evaluate-sure}
  \For {$v \in \winreg$}\label{algline:evaluate-update-begin}
  \State $f(v) \gets h(v)$
  \EndFor\label{algline:evaluate-update-end}
  \EndIf\label{algline:evaluate-no-profitable-switches-end}
  \Until {$f = g$}\label{algline:evaluate-terminate}
  \State \Return $f$
  \EndFunction
\end{algorithmic}
\end{algorithm*}

\subsection{Initialisation}
\label{ssec:initialisation}

\begin{enumerate}
\item
	Determine the almost sure winning region $\winreg$ (and a winning strategy for it) for Player~$0$ (Line~\ref{algline:init-almost}).
	If the winning region is empty, i.e.\ $\winreg = \emptyset$, return (Line~\ref{algline:init-empty-return-begin}-\ref{algline:init-empty-return-end}).

\item
	Solve the remaining game as reachability game with the reachability objective to reach the winning region $\winreg$ (Line~\ref{algline:init-reach}), obtaining the value $\val$ and the strategy $g$ for the vertices with non-zero value.
   \newline
   Improve the strategy according to the reachability computation (Line~\ref{algline:init-improve-begin}-\ref{algline:init-improve-end}).

\item
	Call this algorithm recursively for the sub-game that contains only the states with reachability probability $0$, improving the strategy using results from the recursive calls (Line~\ref{algline:init-recurse-begin}-\ref{algline:init-recurse-end}).

\end{enumerate}

The main task of the initialisation phase performed by $\Call{Initialize}{\mg}$ is to provide a strategy $f$ for Player~$0$ under which the winning probability $\val^{\mg}_{f}(v)$ lies in $]0,1[$ for each vertex $v$ having the optimal winning probability $\val^{\mg}(v) $in $]0,1[$ while
the strategy is winning in the region $\winreg$ where Player~$0$ wins almost surely.
We call such strategies \emph{realising}.

Note that the correctness of the algorithm does not rely on using realising strategies, it is merely a heuristic.
It is chosen to avoid that the algorithm gets stuck by a too large initial winning region of player $1$.

\begin{example}
  \label{exa:initialise}
  We apply \Call{Initialise}{$\functionDot$} on the MPG $\mg_{x}$ from Figure~\ref{fig:example-extended}.
  The most significant steps of the algorithm on $\mg_{x}$ are depicted in the pictures shown in Algorithm~\ref{alg:main}.
  The winning region from Line~\ref{algline:init-almost} is $\winreg = \setnocond{v_{w}}$, depicted as a dashed box.
  Because $v_{0}$ and $v_{3}$ are outside the winning region, the choice of their edges is arbitrary and we can assume that \Call{QualiSolve}{$\functionDot$} returns a strategy in which the edges $e_{0,2}$ and $e_{3,1}$ are chosen, depicted as full edges.
  The dashed edges are those under the control of Player~$1$;
  the dash-dotted edges are those randomly taken.
  As the winning region is nonempty, we do not return in Lines~\ref{algline:init-empty-return-begin}-\ref{algline:init-empty-return-end}.
  The reachability computation in Line~\ref{algline:init-reach} and the following updates now set the choice for $v_{0}$ to $e_{0,1}$;
  regarding $v_{3}$, since initially the value for $v_{3}$ is $0$, the initial choice for $v_{2}$ is $e_{2,2}$, so for $v_{3}$ the choice between $e_{3,1}$ and $e_{3,2}$ is irrelevant.
  After the updates, the only nodes with winning probability $0$ are $v_{l}$, $v_{2}$, and $v_{3}$;
  this means that, in Line~\ref{algline:init-call-recurse}, we call the function recursively with the game consisting of the nodes $\setnocond{v_{l}, v_{2}, v_{3}}$.
  The winning region is now $\winreg = \setnocond{v_{2}, v_{3}}$, obtained by Player~$0$ by choosing $e_{3,2}$, so the following reachability computation in Line~\ref{algline:init-reach} and updates maintain the choice for $v_{3}$ to $e_{3,2}$.
  The only state still with value $0$ is $v_{l}$, so the recursive call has as argument a game consisting only of the node $v_{l}$.
  However, this recursive step is already left at Line~\ref{algline:init-empty-return-mid} because the winning region is empty.
  As there are no Player~$0$ nodes in such a one-node game, the recursive call does not lead to further updates of the strategy.
\end{example}

\subsection{Strategy improvement step}
\label{ssec:step}

Input is a strategy and a
parity game.
Output is a superior strategy and a
parity game -- or an optimality result for the given strategy.

\begin{enumerate}
 \item take a strategy $f$ for Player~$0$ (Line~\ref{algline:improve-head})
 \item construct the parity MDP for it (Line~\ref{algline:construct-mdp})
 \item evaluate the parity MDP (Line~\ref{algline:evaluate-mdp})
 \item if there are profitable switches: choose \& return update among them (Lines~\ref{algline:evaluate-profitable-switches-begin}-\ref{algline:evaluate-profitable-switches-end})
 \newline else (Lines~\ref{algline:evaluate-no-profitable-switches-begin}-\ref{algline:evaluate-no-profitable-switches-end})

\begin{enumerate}
 \item build the sub-game that only uses neutral edges (Line~\ref{algline:evaluate-neutral})
 \item determine almost sure winning region \& strategy (Line~\ref{algline:evaluate-sure})%
 \footnote{optional: for the rest: determine maximal reachability strategy to this region where the region can be reached with probability $>0$}
 on it for Player~$0$
\item if the region is not empty, update the strategy accordingly.
  That is, in this winning region, update the strategy such that it is winning.
  (Lines~\ref{algline:evaluate-update-begin}-\ref{algline:evaluate-update-end})
 \item if the region is empty, terminate ($f$ is optimal) (Line~\ref{algline:evaluate-terminate})
\end{enumerate}

\end{enumerate}

This step is repeated until $f$ is found to be optimal by the algorithm.

\begin{example}
  \label{exa:improve}
  We reconsider the MPG $\mg_{x}$ from Figure~\ref{fig:example-extended} and the strategy $f$ from Example~\ref{exa:initialise} with $f(v_{0}) = e_{0,1}$.
  The evaluation of the induced MDP in Line~\ref{algline:evaluate-mdp} leads to a value of $0.55$ in $v_{0}$ and $v_{1}$.
  There are no profitable switches, so Lines~\ref{algline:evaluate-profitable-switches-begin}-\ref{algline:evaluate-profitable-switches-end} do not lead to any changes of $f$.
  The subgame $\mg'$ computed in Line~\ref{algline:evaluate-neutral} does not contain $e_{1,1}$, because choosing this edge would lead to a value of $0.95$, which is worse than $0.55$ for Player~$1$.
  Evaluating $\mg'$ in Line~\ref{algline:evaluate-sure}, we see that $v_{0}$ and $v_{1}$ are now part of the winning region, because Player~$1$ cannot leave it using $e_{1,1}$.
  The choice for $v_{0}$ is thus updated to $e_{0,2}$.
  In the next iteration of the improvement loop, there are neither profitable switches, nor does the subgame of neutral edges lead to any improvement.
  Thus, the algorithm terminates at this point.
\end{example}

\section{Linear programming for solving reachability Markov decision processes}
\label{apx:reach-lp}
Consider a reachability MDP
$\mg = (\vertexSet_{0}, \emptyset, \vertexSet_{r}, \edgeSet, \prob, \reach)$.
Then we have that, for each $v \in \vertexSet$, we have $\val^{\mg}(v) = w_{v}$, where $w$ is the solution vector obtained from the following linear programming problem:
\[
\begin{array}{lr}
\text{minimise} 
 
\sum \setcond{w_{v}}{v \in \vertexSet} \\ 
\text{subject to} \\
w_{v} \geq 0
&
\forall v \in \vertexSet
\\
w_{v} \leq 1
&
\forall v \in \reach
\\
w_{v} \geq w_{v'}
&
\kern-15mm\forall v \in \vertexSet_{0}, (v,v') \in \edgeSet
\\
w_{v} \geq \sum \setcond{\prob(v)(v') \cdot w_{v'}}{(v,v') \in \edgeSet}
&
\forall v \in \vertexSet_{r}
\end{array}
\]

\section{Reduction from parity to reachability games}
\label{apx:parity2reach}
The reduction from parity to reachability games from \cite{DBLP:conf/isaac/AnderssonM09} focuses on the tractability of the reductions.
This has left us in a tight spot between re-doing a simplified version of the construction -- which is arguably not required -- and referring to a complicated construction that consists of many steps and that does not provide a theorem, which directly makes the claim we make in Theorem \ref{theo:epsilon}.
For this reason, we provide a reduction below.
Note that we make no claim regarding tractability.
This is not required, as the resulting game only occurs in proofs, but is not used in our algorithm.

We use the translation from $\mg$ to $\mg_{\epsilon,n}$ with
\[
	\lprob(\epsilon,n,c) = 
	\begin{cases}
		0 & \text{if $c$ is even,}\\
		\delta^{c+1} & \text{if $c$ is odd}
	\end{cases}
\]
and
\[
	\wprob(\epsilon,n,c) = 
	\begin{cases}
		\delta^{c+1} & \text{if $c$ is even,}\\
		0 & \text{if $c$ is odd}
	\end{cases}
\]
where a suitable $\delta \in (0,1]$ exists with the properties we require for given $n$ and $\epsilon$. (Details follows.)

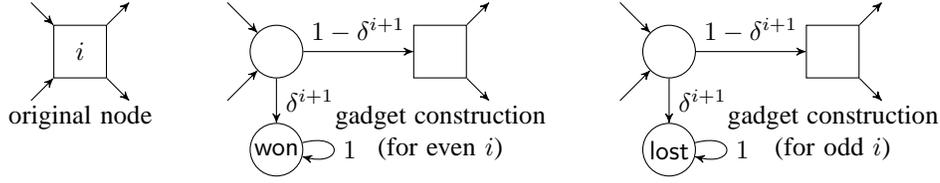
\begin{figure*}
  \centering
  \begin{tikzpicture}[->,>=stealth',auto,scale=0.87]
    \node[p0node] (orig) at (0,0) {$i$};
    \node at ($(orig)+(0,-1)$) {original node};
    
    \draw ($(orig)+(-0.75,-0.75)$) to (orig);
    \draw ($(orig)+(-0.75,0.75)$) to (orig);
    \draw (orig) to ($(orig)+(0.75,-0.75)$);
    \draw (orig) to ($(orig)+(0.75,0.75)$);
    
    \node[prnode] (gadgetEven) at (3,0) {};
    \node[p0node] (origgadgetEven) at ($(gadgetEven)+(2.5,0)$) {};
    \node[prnode] (sink) at ($(gadgetEven)+(0,-1.5)$) {$\won$};
    \node at ($(origgadgetEven)+(0,-1.25)$) {\begin{minipage}{40mm}\centering gadget construction\\(for even $i$)\end{minipage}};

    \draw ($(gadgetEven)+(-0.75,-0.75)$) to (gadgetEven);
    \draw ($(gadgetEven)+(-0.75,0.75)$) to (gadgetEven);
    \draw (gadgetEven) to node[above] {$1-\delta^{i+1}$}  (origgadgetEven);
    \draw (gadgetEven) to node[right] {$\delta^{i+1}$} (sink);
    \draw (origgadgetEven) to ($(origgadgetEven)+(0.75,-0.75)$);
    \draw (origgadgetEven) to ($(origgadgetEven)+(0.75,0.75)$);
    \draw (sink) to [out=375,in=345,looseness=8] node[right] {$1$} (sink);

    \node[prnode] (gadgetOdd) at (9,0) {};
    \node[p0node] (origgadgetOdd) at ($(gadgetOdd)+(2.5,0)$) {};
    \node[prnode] (sink) at ($(gadgetOdd)+(0,-1.5)$) {$\lost$};
    \node at ($(origgadgetOdd)+(0,-1.25)$) {\begin{minipage}{40mm}\centering gadget construction\\(for odd $i$)\end{minipage}};

    \draw ($(gadgetOdd)+(-0.75,-0.75)$) to (gadgetOdd);
    \draw ($(gadgetOdd)+(-0.75,0.75)$) to (gadgetOdd);
    \draw (gadgetOdd) to node[above] {$1-\delta^{i+1}$}  (origgadgetOdd);
    \draw (gadgetOdd) to node[right] {$\delta^{i+1}$} (sink);
    \draw (origgadgetOdd) to ($(origgadgetOdd)+(0.75,-0.75)$);
    \draw (origgadgetOdd) to ($(origgadgetOdd)+(0.75,0.75)$);
    \draw (sink) to [out=375,in=345,looseness=8] node[right] {$1$} (sink);
  \end{tikzpicture}
  \caption{Gadget construction.}
  \label{fig:oldgadget}
\end{figure*}

That is, we obtain the gadget construction from Figure \ref{fig:oldgadget}.
We refer to the translation as $\mg^\delta$.
The main observation when looking at $\delta>0$ is to follow what happens if we let $\delta$ shrink towards $0$.

Let $f_{0}$ and $f_{1}$ be strategies of Player $0$ and $1$ for $\mg$ and $f'_{0}$ and $f'_{1}$ their similar strategies for $\mg^\delta$ (defined in the same way as Theorem \ref{theo:epsilon}).
Let $L \subseteq V$ be a leaf component (a strongly connected component without outgoing edges) in $\mg_{f_{0},f_{1}}$.
Recall that runs almost surely reach (and then get trapped) in some leaf component for Markov chains.
For Markov chains with a parity acceptance condition, the runs that reach a leaf component $L$ are almost surely accepting if the minimal priority $c_L = \min \setcond{\pri(v)}{v \in L}$ of the states in $L$ is even, and they are almost surely losing if $c_L$ is odd.

We make the following simple observations for $\mg^\delta_{f'_{0},f'_{1}}$.
\begin{enumerate}
\item[(1)] If $\delta$ goes to $0$, the chance of reaching states in $L' = L \cup \setcond{v'}{v \in L}$ in $\mg^\delta_{f'_{0},f'_{1}}$ from a state $v_0 \in V$ goes to the chance of reaching $L$ in $\mg_{f_{0},f_{1}}$ from $v_0$.
\item[(2)] When starting in $L'$, there is a $\delta' \in (0,0.5)$ such that, for all $\delta \in (0,\delta')$, the chance of leaving to $\won$ or $\lost$ before visiting a vertex $v'$ (the primed copy of $v$) with $\pri(v) = c_L$ is smaller than $\delta^{c+1.5}$.
\item[(3)] When starting in $L'$ where $c_L$ is even, there is a $\delta'>0$ such that, for all $\delta \in (0,\delta')$, the chance of reaching $\lost$ is at most $\sqrt[3]{\delta}$ times the chance of reaching $\won$.
\item[(4)] When starting in $L'$ where $c_L$ is odd, there is a $\delta'>0$ such that, for all $\delta \in (0,\delta')$, the chance of reaching $\won$ is at most $\sqrt[3]{\delta}$ times the chance of reaching $\lost$.
\item[(5)] When $\delta$ goes to $0$, the chance of reaching the $\won$ from a state $v_0 \in V$ goes to the chance of winning in $\mg_{f_{0},f_{1}}$ from $v_0$.
\end{enumerate}

To see (1), if we want to approach the likelihood with precision $2\epsilon$, we choose an $n$ such that $L'$ is reached after more than $n$ states with chance $<n$, and then choose $\delta\in (0,1)$ such that $(1-\delta)^n > 1-\epsilon$. The latter property provides that the difference in the chance of reaching $L$ in $n$ steps and reaching $L'$ in $2n$ steps is less than $\epsilon$.

To see (2), consider that, as $L$ is a leaf component, there is positive probability from every vertex $w \in L$ to reach a vertex $v$ with minimal $\pri(v)=c_L$ within $n=|L|$ steps in $\mg_{f,g}$.
Let $p_{\min} > 0$ be the smallest such probability. 
Then, in $L'$, a vertex $v'$, which is the primed copy of a vertex $v$ with minimal $\pri(v)=c_L$, can be reached within $2n$ steps in $\mg^\delta_{f',g'}$ with chance at least $p_{\min}(1-\delta^{c+2})^n > \frac{p_{\min}}{2^n}$.
The chance to reach $\won$ or $\lost$ within $n$ steps and without reaching such a vertex $v'$ first is at most $n\cdot \delta^{c+2}$.

Consequently, the chance of reaching $\won$ or $\lost$ prior to reaching a vertex $v'$ which is the primed copy of a vertex $v$ with minimal colour $\pri(v)=c_L$ is 
at most $\frac{n \cdot \delta^{c+2}}{n\cdot \delta^{c+2} + \frac{p_{\min}}{2^n}}$. If we choose $\delta$ small enough that $\delta^{c+2} < \frac{p_{\min}}{n\cdot 2^n}$ and $\sqrt{\delta}< \frac{2p_{\min}}{n \cdot 2^n}$, then we get
\[\frac{n \cdot \delta^{c+2}}{n\cdot \delta^{c+2} + \frac{p_{\min}}{2^n}} < \frac{n \cdot 2^n \delta^{c+2}}{2p_{\min}}< \delta^{c+1.5},\]
which provides the claim.

(3) and (4) follow immediately when $\delta$ is small enough such that $\frac{\delta^{c+1.5}}{\delta^{c+1}(1-\delta^{c+1.5})} = \frac{\sqrt{\delta}}{(1-\delta^{c+1.5})} < \sqrt[3]{\delta}$ holds.

(5) finally follows with the previous points and the observation that $\won$ and $\lost$ are the only leaf components in $\mg^\delta_{f'_{0},f'_{1}}$, and are therefore reached almost surely.

This establishes all properties we need for Theorem \ref{theo:epsilon}.
\fi

\end{document}